\documentstyle[12pt]{article}
\def\journal{\topmargin .3in	\oddsidemargin .5in
	\headheight 0pt	\headsep 0pt
	\textwidth 5.625in 
	\textheight 8.25in 
	\marginparwidth 1.5in
	\parindent 2em
	\parskip .5ex plus .1ex		\jot = 1.5ex}
\journal

\catcode`\@=11
\def\marginnote#1{}
\newcount\hour
\newcount\minute
\newtoks\amorpm
\hour=\time\divide\hour by60
\minute=\time{\multiply\hour by60 \global\advance\minute by-\hour}
\edef\standardtime{{\ifnum\hour<12 \global\amorpm={am}%
	\else\global\amorpm={pm}\advance\hour by-12 \fi
	\ifnum\hour=0 \hour=12 \fi
	\number\hour:\ifnum\minute<10 0\fi\number\minute\the\amorpm}}
\edef\militarytime{\number\hour:\ifnum\minute<10 0\fi\number\minute}
\def\draftlabel#1{{\@bsphack\if@filesw {\let\thepage\relax
   \xdef\@gtempa{\write\@auxout{\string
      \newlabel{#1}{{\@currentlabel}{\thepage}}}}}\@gtempa
   \if@nobreak \ifvmode\nobreak\fi\fi\fi\@esphack}
	\gdef\@eqnlabel{#1}}
\def\@eqnlabel{}
\def\@vacuum{}
\def\draftmarginnote#1{\marginpar{\raggedright\scriptsize\tt#1}}
\def\draft{\oddsidemargin -.5truein
	\def\@oddfoot{\sl preliminary draft \hfil
	\rm\thepage\hfil\sl\today\quad\militarytime}
	\let\@evenfoot\@oddfoot	\overfullrule 3pt
	\let\label=\draftlabel
	\let\marginnote=\draftmarginnote
   \def\@eqnnum{(\theequation)\rlap{\kern\marginparsep\tt\@eqnlabel}%
\global\let\@eqnlabel\@vacuum}  }
\def\preprint{\twocolumn\sloppy\flushbottom\parindent 2em
	\leftmargini 2em\leftmarginv .5em\leftmarginvi .5em
	\oddsidemargin -.5in	\evensidemargin -.5in
	\columnsep .4in	\footheight 0pt
	\textwidth 10in	\topmargin  -.4in
	\headheight 12pt \topskip .4in
	\textheight 7.1in \footskip 0pt
	\def\@oddhead{\thepage\hfil\addtocounter{page}{1}\thepage}
	\let\@evenhead\@oddhead	\def\@oddfoot{}	\def\@evenfoot{} }
\def\numberbysection{\@addtoreset{equation}{section}
	\def\theequation{\thesection.\arabic{equation}}}
\def\underline#1{\relax\ifmmode\@@underline#1\else
	$\@@underline{\hbox{#1}}$\relax\fi}
\def\titlepage{\@restonecolfalse\if@twocolumn\@restonecoltrue\onecolumn
     \else \newpage \fi \thispagestyle{empty}\c@page\z@
	\def\thefootnote{\fnsymbol{footnote}} }
\def\endtitlepage{\if@restonecol\twocolumn \else \newpage \fi
	\def\thefootnote{\arabic{footnote}}
	\setcounter{footnote}{0}}  
\catcode`@=12
\relax
\def\figcap{\section*{Figure Captions\markboth
	{FIGURECAPTIONS}{FIGURECAPTIONS}}\list
	{Figure \arabic{enumi}:\hfill}{\settowidth\labelwidth{Figure 999:}
	\leftmargin\labelwidth
	\advance\leftmargin\labelsep\usecounter{enumi}}}
 \relax
\def\tablecap{\section*{Table Captions\markboth
	{TABLECAPTIONS}{TABLECAPTIONS}}\list
	{Table \arabic{enumi}:\hfill}{\settowidth\labelwidth{Table 999:}
	\leftmargin\labelwidth
	\advance\leftmargin\labelsep\usecounter{enumi}}}
 \relax
\def\reflist{\section*{References\markboth
	{REFLIST}{REFLIST}}\list
	{[\arabic{enumi}]\hfill}{\settowidth\labelwidth{[999]}
	\leftmargin\labelwidth
	\advance\leftmargin\labelsep\usecounter{enumi}}}
 \relax
\makeatletter
\newcounter{pubctr}
\def\publist{\@ifnextchar[{\@publist}{\@@publist}}
\def\@publist[#1]{\list
	{[\arabic{pubctr}]\hfill}{\settowidth\labelwidth{[999]}
	\leftmargin\labelwidth
	\advance\leftmargin\labelsep
	\@nmbrlisttrue\def\@listctr{pubctr}
	\setcounter{pubctr}{#1}\addtocounter{pubctr}{-1}}}
\def\@@publist{\list
	{[\arabic{pubctr}]\hfill}{\settowidth\labelwidth{[999]}
	\leftmargin\labelwidth
	\advance\leftmargin\labelsep
	\@nmbrlisttrue\def\@listctr{pubctr}}}
 \relax
\makeatother
\catcode`\@=11
\def\section{\@startsection {section}{1}{0pt}{-3.5ex plus -1ex minus
 -.2ex}{2.3ex plus .2ex}{\raggedright\large\bf}}
\catcode`\@=12
\newskip\humongous \humongous=0pt plus 1000pt minus 1000pt
\def\caja{\mathsurround=0pt}

\newif\ifdtup
\def\panorama{\global\dtuptrue \openup1\jot \caja
	\everycr{\noalign{\ifdtup \global\dtupfalse
	\vskip-\lineskiplimit \vskip\normallineskiplimit
	\else \penalty\interdisplaylinepenalty \fi}}}
\def\eqalignno#1{\panorama \tabskip=\humongous
	\halign to\displaywidth{\hfil$\displaystyle{##}$
	\tabskip=0pt&$\displaystyle{{}##}$\hfil
	\tabskip=\humongous&\llap{$##$}\tabskip=0pt
	\crcr#1\crcr}}

\def\oldreffmt#1{\rlap{[#1]} \hbox to 2\parindent{}}

\def\figfmt#1{\rlap{Figure {#1}} \hbox to 1in{}}

\def\abs#1{\left| #1\right|}

\def\half{{1\over 2}}
\def\beq{\begin{equation}}
\def\eeq{\end{equation}}

\def\bea{\begin{eqnarray}}

\def\eea{\end{eqnarray}}
\catcode`\@=11
\def\eqnarray{\stepcounter{equation}\let\@currentlabel=\theequation
\global\@eqnswtrue
\global\@eqcnt\z@\tabskip\@centering\let\\=\@eqncr
\gdef\@@fix{}\def\eqno##1{\gdef\@@fix{##1}}%
$$\halign to \displaywidth\bgroup\@eqnsel\hskip\@centering
  $\displaystyle\tabskip\z@{##}$&\global\@eqcnt\@ne
  \hskip 2\arraycolsep \hfil${##}$\hfil
  &\global\@eqcnt\tw@ \hskip 2\arraycolsep $\displaystyle\tabskip\z@{##}$\hfil
   \tabskip\@centering&\llap{##}\tabskip\z@\cr}

\def\@@eqncr{\let\@tempa\relax
    \ifcase\@eqcnt \def\@tempa{& & &}\or \def\@tempa{& &}
      \else \def\@tempa{&}\fi
     \@tempa \if@eqnsw\@eqnnum\stepcounter{equation}\else\@@fix\gdef\@@fix{}\fi
     \global\@eqnswtrue\global\@eqcnt\z@\cr}

\catcode`\@=12
\font\tenbifull=cmmib10 
\font\tenbimed=cmmib10 scaled 800
\font\tenbismall=cmmib10 scaled 666
\def\boldzeta{\fam=9{\mathchar"7110 } }

\def\boldlambda{\fam=9{\mathchar"7115 } }

\relax

\def\
thefootnote{\fnsymbol{footnote}}
\def\wt{\widetilde}
\def\wh{\widehat}
\renewcommand{\boldlambda}{\mbox{\boldmath $\lambda$}}
\renewcommand{\boldzeta}{\mbox{\boldmath $\zeta$}}
\begin{document}
\begin{titlepage}
\begin{center}
\today          \hfill LBL-37343\\
\hfill UCB-PTH-95/17\\
\hfill hep-ph/9508288 \\
\vskip .1in

\vskip .25in

{\large \bf Flavor Mixing Signals For Realistic\\  Supersymmetric Unification.}
\footnote{This work was supported in part by the Director, Office of
Energy Research, Office of High Energy and Nuclear Physics, Division of
High Energy Physics of the U.S. Department of Energy under Contract
DE-AC03-76SF00098 and in part by the National Science Foundation under
grant PHY-90-21139.}

\vskip .25in
{\bf Nima Arkani-Hamed\\
Hsin-Chia Cheng}\\
and\\
{\bf L.J. Hall}\\
\vskip .20in

{\em Theoretical Physics Group\\
    Lawrence Berkeley Laboratory\\
and\\
Department of Physics\\
      University of California\\
    Berkeley, California 94720}
\end{center}

\vskip .25in


\begin{abstract}
The gauge interactions of any supersymmetric extension of the standard
model involve new flavor mixing matrices.
The assumptions involved in the construction of minimal supersymmetric models,
both $SU(3) \times SU(2) \times U(1)$ and
grand unified theories, force a large degree of triviality on these matrices.
However, the requirement of realistic quark and lepton masses in supersymmetric
grand unified theories forces these matrices to be non-trivial.
This leads to important new dominant contributions to the
neutron electric dipole moment
and to the decay mode $p \to K^o\mu^+$, and suggests that there may be
important
weak scale radiative corrections to the Yukawa coupling matrix of the up
quarks.
The lepton flavor violating signal $\mu \to e\gamma$ is studied in these
theories when $\tan\beta$ is sufficiently large that radiative
effects of couplings
other than
$\lambda_t$ must be included.
The naive expectation that large $\tan\beta$ will force sleptons to
unacceptably large masses is not
borne out: radiative suppressions to the leptonic flavor mixing angles
allow regions where the
sleptons are as light as 300 GeV, provided the top Yukawa
coupling in the unified
theory is near the minimal value consistent with $m_t$.

\end{abstract}

\end{titlepage}

\renewcommand{\thepage}{\roman{page}}
\setcounter{page}{2}
\mbox{ }

\vskip 1in

\begin{center}
{\bf Disclaimer}
\end{center}

\vskip .2in

\begin{scriptsize}
\begin{quotation}
This document was prepared as an account of work sponsored by the United
States Government. While this document is believed to contain correct
 information, neither the United States Government nor any agency
thereof, nor The Regents of the University of California, nor any of their
employees, makes any warranty, express or implied, or assumes any legal
liability or responsibility for the accuracy, completeness, or usefulness
of any information, apparatus, product, or process disclosed, or represents
that its use would not infringe privately owned rights.  Reference herein
to any specific commercial products process, or service by its trade name,
trademark, manufacturer, or otherwise, does not necessarily constitute or
imply its endorsement, recommendation, or favoring by the United States
Government or any agency thereof, or The Regents of the University of
California.  The views and opinions of authors expressed herein do not
necessarily state or reflect those of the United States Government or any
agency thereof of The Regents of the University of California and shall
not be used for advertising or product endorsement purposes.
\end{quotation}
\end{scriptsize}

\vskip 2in

\begin{center}
\begin{small}
{\it Lawrence Berkeley Laboratory is an equal opportunity employer.}
\end{small}
\end{center}

\newpage
\renewcommand{\thepage}{\arabic{page}}
\setcounter{page}{1}

\noindent{\bf I. Introduction}

\medskip

It has recently been demonstrated that flavor and CP
violation provide an important new probe of supersymmetric
grand unified theories [1-4].
These new signals, such as $\mu\to e\gamma$ and the electron electric
dipole moment $d_e$, are complementary to the classic tests of
proton decay, neutrino masses
and quark and charged lepton mass relations.
The classic tests are very dependent on the flavor interactions
and symmetry breaking sector of the unified model: it is only too
easy to construct models in which these signals are absent or unobservable.
However, they are insensitive to the hardness scale, $\Lambda_H$,
of supersymmetry breaking.\footnote{This is the highest scale at
which supersymmetry breaking squark and gluino masses appear
in the theory as local interactions.} On the other hand, the new flavor and
CP violating signals are relatively insensitive
to the form of the flavor interactions and unified gauge symmetry
breaking, but are absent if the hardness scale, $\Lambda_H$,
 falls beneath the unified
scale, $M_G$. The signals are generated by the unified flavor
interactions leaving
an imprint on the form of the soft supersymmetry breaking operators
\cite{HKR}, which is only possible if supersymmetry
 breaking is present in the unified theory at scales above $M_G$.

The flavor and CP violating signals have been computed in the minimal
$SU(5)$ and $SO(10)$
models for leptonic [1-3] and hadronic processes \cite{BHS2},
for moderate values of tan$\beta$, the ratio of the two Higgs
vacuum expectation values.
While rare muon decays provide an important probe of $SU(5)$,
it is the $SO(10)$ theory which is
most powerfully tested.
If the hardness scale for supersymmetry breaking is large enough, as in the
popular
supergravity models, it may be possible for the minimal $SO(10)$ theory to
be probed throughout the interesting
range of superpartner masses by searches for $\mu \to e\gamma$ and $d_e$.

The flavor changing and CP violating probes of $SO(10)$ are sufficiently
powerful
to warrant an exploration of consequences for non-minimal models,
which is the subject of this
paper. In particular, we study $SO(10)$ theories in which

(I) {\it{The Yukawa interactions are non-minimal.}}

In the minimal model the quarks and leptons lie in three
 16's and the two Higgs doublets $H_U$ and
$H_D$ lie in two 10 dimensional
representations $10_U$ and $10_D$.
The quark and charged lepton masses are assumed to arise from the interactions
$16 {\boldlambda}_U 16\ 10_U + 16 {\boldlambda}_D 16\ 10_D$.
This model is a useful fiction: it is very simple to work with, but leads to
the
mass relation $m_e/m_\mu = m_d/m_s$,
which is in error by an order of magnitude.
It is clearly necessary to introduce a mechanism to insert $SO(10)$
breaking into the Yukawa interactions.
The simplest way to achieve this is to assume that at the unification scale,
 $M_G$,
some of the Yukawa interactions arise from higher dimensional
operators
involving fields $A$ which break the $SO(10)$ symmetry group.
This implies that ${\boldlambda}_{U,D} \to {\boldlambda}_{U,D}(A)$.
Every realistic model of $SO(10)$ which has been constructed
has this form; hence one
should view this generalization of the minimal model as a necessity.

(II) {\it{The ratio of electroweak VEV's,}} $\tan\beta = v_U/v_D$,
{\it{ is allowed to be large,}}
$\approx m_t/m_b$.

This is certainly not a necessity; to the contrary,
a simple extrapolation of the results
of \cite{BHS} to such large values of $\tan\beta$ suggests that it is already
excluded by the present limit on $\mu \to e\gamma$.
The case of large $\tan\beta$ in $SO(10)$ has received much attention
\cite{ALS,HRS,RS,ADHRS} partly because it has important ramifications for the
origin of $m_t/m_b = (\lambda_t/\lambda_b) \tan\beta$.
To what extent is this puzzling large ratio to be understood as a large
hierarchy
of Yukawa couplings, and to what extent in terms
of a large value for $\tan\beta$?
If the third generation masses arise from a single interaction of the
form $16_3 16_3 10$
it is possible to predict $m_t$ using $m_b$ and $m_\tau$ as input
\cite{ALS}, providing the theory is
perturbative up to $M_G$.
The prediction is $175 \pm 10 $ GeV \cite{HRS}, and requires
$\tan\beta \approx m_t/m_b$.
In this paper we investigate whether this intriguing possibility is
excluded by the
$\mu \to e \gamma$ signal; or, more correctly, we determine whether
it requires a soft
origin for supersymmmetry breaking, making it
incompatible with the standard supergravity scenario \cite{CAN}.

In the next section we show that $SO(10)$ models with ${\boldlambda} \to
{\boldlambda} (A)$
possess new gaugino mixing matrices in the up-quark sector, which did
not arise in the minimal models.
In section III we set our notation for the supersymmetric standard model with
arbitrary gaugino mixing matrices, and we show which mixing matrices
are expected from unified models according to the gauge group
and the value of $\tan\beta$.
In section IV we describe the new phenomenological signatures
which are generated by the gaugino mixing
matrices in the up sector; these signatures are generic to
all models with Yukawa interactions
generated from higher dimensional operators.
The consequences of large $\tan\beta$ for the flavor and CP
violating signatures are analyzed analytically  in section V
and numerically in section VI.
The analysis of the first five sections applies to a wide class of models.
In section VII we illustrate the results in the particular
models introduced by Anderson et al \cite{ADHRS}.
As well as providing illustrations, these models have features
unique to themselves.
Conclusions are drawn in section VIII.

\newpage

\noindent{\bf II. New Flavor Mixing in the Up Sector}

\medskip

In [1-4] flavor and CP violating signals are studied in
minimal $SU(5)$ and $SO(10)$ models with moderate $\tan\beta$.
In these models the
radiative corrections
to the scalar mass matrices are dominated by the top quark
Yukawa coupling $\lambda_t$ of the
unified theory, so the scalar mass matrices tend to align with the
up-type Yukawa coupling matrix and all non-trivial flavor mixing matrices
are simply related to the KM matrix.
However, as mentioned above, the minimal models do not give realistic
fermion masses.
One has to insert $SO(10)$ breaking into the Yukawa interactions.
The simplest way to achieve this is to assume that the light fermion masses
come from the non-renormalizable operators
$$
\lambda'_{ij} 16_i{A_1\over M_1}{A_2\over M_2}\ldots {A_\ell\over M_\ell}
10 {A_{\ell+1}\over M_{\ell +1}} \ldots {A_n\over M_{n}}  16_j,\eqno(2.1)
$$
where the $16_i$'s contain the three low energy families, 10
contains the Higgs doublets, and $A$'s are adjoint fields with
vacuum expectation values (VEV's)
which break the $SO(10)$ gauge group.
After substituting in the VEV's of the adjoints, they
become the usual Yukawa interactions with different Clebsch factors
associated with Yukawa couplings of fields with different quantum numbers.
For example in the models introduced by Anderson et al. \cite{ADHRS},
(hereafter
referred to as ADHRS models)
$$
{\boldlambda}_U = \pmatrix{0&z_uC &0\cr
                z'_uC & y_uE &x_uB\cr
                0 & x'_uB &A},
{\boldlambda}_D = \pmatrix{ 0 & z_dC & C\cr
                   z'_d & y_dE & x_dB\cr
                    0 & x'_dB &A},
{\boldlambda}_E = \pmatrix{ 0 & z_eC & 0\cr
                  z'_eC & y_eE & x_eB\cr
                  0 & x'_eB & A},\eqno(2.2)
$$
where the $x,y,z$'s are Clebsch factors arising from the VEV's
of the adjoint fields.
Thus realistic fermion masses and mixings can be obtained.

The radiative corrections to the soft SUSY-breaking operators
above $M_G$ are now more complicated.
{}From the interactions (2.1) the following soft supersymmetry
breaking operators are generated:
$$
\lambda^\dagger_{ik}(A) m^2_{k\ell}(A) \lambda_{\ell j}(A) \;
\phi^\dagger_i\phi_j,\eqno(2.3)
$$
where $\phi_i, \phi_j$ are scalar components of the superfields, and
$\lambda_{ij}(A)$ are adjoint dependent couplings,
$\lambda(A) = \lambda' {A_1\over M_1} \ldots {A_n\over M_n}$.
After the adjoints take their VEV's, the
$m^2_{k\ell} (A)$  become the usual soft scalar masses.
If we ignore the wavefunction renormalization of the adjoint fields
(which is valid in the one-loop approximation), this is the same as if
we had replaced the adjoints by their VEV's all the
way up to the ultraheavy scale where the ultraheavy fields are integrated out,
 and treated these nonrenormalizable operators as the usual Yukawa
interactions and scalar mass operators.
This is a convenient way of thinking and we will use it in the rest
of the paper.

Above the GUT scale, in addition to the Yukawa interactions
which give the fermion masses
$$
Q{\boldlambda}_U U^cH_U, \; Q{\boldlambda}_D D^c H_D, \; E^c
{\boldlambda}_E L H_D, \eqno(2.4)
$$
the operators (2.1) also lead to
$$
\eqalignno{
Q{\boldlambda}_{qq} Q H_{U_3}, &E^c {\boldlambda}_{eu}U^c H_{U_3}, \;
 N{\boldlambda}_{nd}D^cH_{U_3},\cr
Q \lambda_{q\ell} L H_{D_3}, &U^c{\boldlambda}_{ud}D^c H_{D_3},
N{\boldlambda}_{n\ell} LH_U,&(2.5)\cr}
$$
where $H_{U_3}, H_{D_3}$ are the triplet partners of the two
Higgs doublets $H_U$ and $H_D$.
Each Yukawa matrix has different Clebsch factors associated with
its elements, so they can not be diagonalized in
the same basis.
The scalar mass matrices receive radiative corrections from Yukawa
interactions of both (2.4) and (2.5), which, in the one-loop approximation,
take the form
$$
\eqalignno{
\Delta{\bf{m}}^2_Q &\propto {\boldlambda}_U{\boldlambda}^\dagger_U +
     {\boldlambda}_D{\boldlambda}_D^\dagger + 2{\boldlambda}_{qq}
    {\boldlambda}^\dagger_{qq}+ {\boldlambda}_{q\ell}
    {\boldlambda}^\dagger_{q\ell},\cr
\Delta{\bf{m}}^2_U &\propto 2{\boldlambda}^\dagger_U{\boldlambda}_U +
    {\boldlambda}^\dagger_{eu}{\boldlambda}_{eu} + 2 {\boldlambda}_{ud}
    {\boldlambda}^\dagger_{ud},\cr
\Delta{\bf{m}}^2_D &\propto 2{\boldlambda}^\dagger_D{\boldlambda}_D +
    {\boldlambda}^\dagger_{nd}{\boldlambda}_{nd} + 2{\boldlambda}_{ud}^\dagger
    {\boldlambda}_{ud},\cr
\Delta{\bf{m}}^2_L &\propto {\boldlambda}^\dagger_E{\boldlambda}_E +
3{\boldlambda}^\dagger_{q\ell}  {\boldlambda}_{q\ell}
+ {\boldlambda}^\dagger_{n\ell}{\boldlambda}_{n\ell},\cr
\Delta{\bf{m}}^2_E &\propto 2{\boldlambda}_E{\boldlambda}^\dagger_E
+ 3{\boldlambda}_{eu}{\boldlambda}^\dagger_{eu}.&(2.6)
\cr}
$$
In the minimal $SO(10)$ model, scalar mass renormalizations above $M_G$
arise from a single matrix
${\boldlambda}_U$.
It is therefore possible to choose a ``U-basis''
in which the scalings are purely
diagonal. This is clearly not possible in the general models.
All scalar mass matrices and Yukawa matrices are in general
diagonalized in different bases.
Therefore, flavor mixing matrices should appear in all gaugino vertices,
including in the up-quark
sector (where they are trivial in the minimal models studied in [1-4]).
The up-type quark-squark-gaugino flavor mixing is a novel
feature of the general models.
Its consequences will be discussed in Sec. IV.
Also, the flavor mixing matrices are no longer simply the KM matrix.
They are model dependent and are
 different for different types of quarks and charged leptons,
 and are fully described in the next section.

\newpage

\noindent {\bf III. Flavor Mixing Matrices in General Superymmetric Standard
Models.}

\medskip

In this section we set our notation
for the gaugino flavor mixing matrices in the
supersymmetric theory below $M_G$, taken to have minimal field content.
We also give general expectations for these matrices in a wide variety
of unified theories.

The most general scalar masses are $6 \times 6$
matrices for squarks and charged
sleptons and $3 \times 3$ matrix for sneutrinos,
$$
\eqalignno{
{\bf{m}}^2_{{U}} &=
\pmatrix{{\bf{ m}}^2_{{U}_L} &
({\boldzeta}_U+{\boldlambda}_U\mu\cot \beta)v_U\cr
({\boldzeta}^\dagger_U +
{\boldlambda}^\dagger_U\mu\cot\beta)v_U &
{\bf{m}}^2_{{U}_R} },\cr
   {\bf{m}}^2_{{D}} &=
\pmatrix{ {\bf{m}}^2_{{D}_L} & ({\boldzeta}_D
    +{\boldlambda}_D\mu\tan\beta)v_D\cr
({\boldzeta}^\dagger_D +
{\boldlambda}^\dagger_D \mu\tan\beta) v_D &
    {\bf{m}}^2_{{D}_R}
 },\cr
{\bf{m}}^2_{{E}} &=
\pmatrix{
{\bf{m}}^2_{{E}_L} & ({\boldzeta}_E +
{\boldlambda}_E\mu\tan\beta)v_D\cr
({\boldzeta}^\dagger_E +
{\boldlambda}^\dagger_E\mu\tan\beta)v_D
    & {\bf{m}}^2_{{E}_R} },
\cr
{\bf{m}}^2_{{\nu}} &= \left( {\bf{m}}^2_{{\nu}_{ij}}
\right),&(3.1)\cr}
$$
where ${\bf{m}}^2_{{U}_L}, {\bf{m}}^2_{{D}_L},
 {\bf{m}}^2_{{U}_R},
{\bf{m}}^2_{{D}_R}, {\bf{m}}^2_{{E}_L}, {\bf{m}}^2_{{E}_R}$
 are $3 \times 3$ soft
SUSY-breaking mass matrices for the left-handed and right-handed squarks and
sleptons, and ${\boldzeta}_U, {\boldzeta}_D, {\boldzeta}_E$ are the trilinear
soft SUSY-breaking
terms. To calculate  flavor violating processes,
such as $\mu \to e\gamma$, one
can diagonalize the mass matrix ${\bf{m}}^2_{{E}}$ by the
$6\times 6$ unitary
rotation matrix $V_E$ and ${\bf{m}}^2_{{\nu}}$ by the 3 $\times $ 3 unitary
rotation $V_\nu$,
$$
{\bf{m}}^2_{{E}} = V_E \overline{\bf{m}}^2_{{E}} V^\dagger_E,
 \hskip .25in
{\bf{m}}^2_{{\nu}} = V_\nu \overline{\bf{m}}^2_{{\nu}}
V^\dagger_\nu,\eqno(3.2)
$$
where ${\bf{\overline{m}}}^2_{{E}},
{\bf{\overline{m}}}^2_{{\nu}}$
 are diagonal.
The amplitude for $\mu \to e\gamma$ is given by the
diagrams in Fig.\ 1, summing up all the internal scalar mass eigenstates.

If the entries in the scalar mass matrices are arbitrary, they generally give
unacceptably large rates for flavor violating processes. From the
experimental limits one expects that the first two generation scalar masses
should be approximately degenerate and the chirality-changing mass matrices
${\boldzeta}_A$ should be approximately proportional
to the corresponding Yukawa
coupling matrices ${\boldlambda}_A$.
In this paper we treat the
chirality-conserving mass matrices and chirality-changing mass matrices
separately,
i.e., the mass eigenstates are assumed to be purely left-handed or
right-handed,
and the chirality-changing mass terms are treated as a perturbation.
This may not be a good approximation for the third generation where
the Yukawa
couplings are large,
the correct treatment will be used in the numerical studies of Sec.\ VI.
The superpotential contains
$$
W \supset Q^T {{\boldlambda}}_U U^c H_U  +Q^T
{{\boldlambda}}_D D^c {H}_D +
E^{cT}{{\boldlambda}}_E L {H}_D,\eqno(3.3)
$$
where ${{\boldlambda}}_U, {\boldlambda}_D, {{\boldlambda}}_E$
 are the Yukawa coupling matrices
which are diagonalized by the left and right rotations,
$$
\eqalignno{
{\boldlambda}_U &= V^*_{U_L}\overline{\boldlambda}_U V^\dagger_{U_R},\cr
{\boldlambda}_D &= V^*_{D_L}\overline{\boldlambda}_D V^\dagger_{D_R},\cr
{\boldlambda}_E &= V^*_{E_R}\overline{\boldlambda}_EV^\dagger_{E_L}.&(3.4)\cr}
$$
The soft SUSY-breaking interactions contain
$$
\eqalignno{
\widetilde{Q}^\dagger {\bf{m}}^{2*}_Q \widetilde{Q}
&+ \widetilde{U}^{c\dagger}
{\bf{m}}^2_U
\widetilde{U}^c + \widetilde{D}^{c\dagger} {\bf{m}}^2_D
\widetilde{D}^c + \widetilde{L}^\dagger
{\bf{m}}^2_L \widetilde{L} + \widetilde{E}^{c\dagger}
{\bf{m}}^{2*}_E \widetilde{E}^c\cr
 &+ \widetilde{Q}^T {{\boldzeta}}_U \widetilde{U}^c H_U +
\widetilde{Q}^T {{\boldzeta}}_D
 \widetilde{D}^c {H}_D + \widetilde{E}^{cT}{{\boldzeta}}_E
\widetilde{L}{H}_D.&(3.5)
 \cr}
$$
Because the trilinear terms should be approximately proportional  to the Yukawa
couplings, we write
$$
{{\boldzeta}} = {{\boldzeta}}_0 + \Delta{{\boldzeta}} = A {\boldlambda} +
{{\Delta}}{\boldzeta}.\eqno(3.6)
$$
The soft-breaking mass matrices are diagonalized by:
$$
\eqalignno{
{\bf{m}}^{2*}_Q &= U_Q \overline{\bf{m}}^{2*}_Q U^\dagger_Q, \;\;
{\bf{m}}^2_U     = U_U \overline{\bf{m}}^2_U U^\dagger_U, \;\;
{\bf{m}}^2_D     = U_D\overline{\bf{m}}^2_D U^\dagger_D, \cr
{\bf{m}}^2_L     &= U_L\overline{\bf{m}}^2_L U^\dagger_L, \;\;
{\bf{m}}^{2*}_E = U_E\overline{\bf{m}}^{2*}_E U^\dagger_E ,
&(3.7)\cr}
$$
$$
{{\Delta{\boldzeta}}}_U = V^{'*}_{U_L}\Delta \overline{{\boldzeta}}_U
V^{'\dagger}_{U_R},\;
{{\Delta\boldzeta}}_D = V^{'*}_{D_L} {{\Delta}}\overline{{\boldzeta}}_D
V^{'\dagger}_{D_R},\;
{{\Delta\boldzeta}}_E = V^{'*}_{E_R} \Delta \overline{\boldzeta}_E
V^{'\dagger}_{E_L}.\eqno(3.8)
$$
In the mass eigenstate basis the rotation matrices $V, U$ appear in the gaugino
couplings,
\newpage
$$
\eqalignno{
{\cal{L}}_g &= \sqrt{2} g' \sum^4_{\pi = 1}
 \bigg[ -\half \overline{e}_L
W^\dagger_{E_L}\widetilde{e}_L N_n(H_{n\widetilde{B}}
+ \cot \theta_W H_{n\widetilde{w}_{3}}) +
\overline{e}^c_L W^\dagger_{E_R} \widetilde{e}_R
N_n H_{n\widetilde{B}}\cr
&+ \half \cot \theta_W\overline{\nu}_L\widetilde{\nu}_L N_n
H_{n\widetilde{w}_{3}}\cr
&+\overline{u}_L W^\dagger_{U_L} \widetilde{u}_L N_n({1\over 6}
H_{n\widetilde{B}}+\half\cot\theta_W H_{n\widetilde{w}_3})
+\overline{d}_LW^\dagger_{D_L} \wt{d}_L N_n({1\over 6}
H_{n\widetilde{B}}-\half\cot\theta_WH_{n\widetilde{w}_3})\cr
&- {2\over 3} \overline{u}_L^c W^\dagger_{U_R}
 \widetilde{u}_R N_n H_{n\widetilde{B}} +
{1\over 3} \bar{d}^c_L W^\dagger_{D_R}\widetilde{d}_R
N_n H_{n\widetilde{B}} + h.c.\bigg]\cr
&+ g\sum^2_{c=1} [ \bar{e}_L W^\dagger_{E_L} \widetilde{\nu}_L
(\chi_c K_{c\widetilde{w}})
+\bar{\nu}_L \widetilde{e}_L (\chi^\dagger_c K^*_{c\widetilde{w}})\cr
&+ \bar{d}_L W^\dagger_{D_L} \widetilde{u}_{L} (\chi_c K_{c\widetilde{w}})
 + \bar{u}_L
W^\dagger_{U_L}\widetilde{d}_L
(\chi^\dagger_c K^* _{c\widetilde{w}}) +h.c.]\cr
&+\sqrt{2} g_3 [\bar{u}_L W^\dagger_{U_L}\widetilde{u}_L\widetilde{g}
 + \bar{d}_L
W^\dagger_{D_L} \widetilde{d}_L \widetilde{g}
+ \bar{u}^c_L W^\dagger_{U_R}
\widetilde{u}_R\widetilde{g} +
\overline{d}^c_LW^\dagger_{D_R}\wt{d}_R\wt{g} + h.c.],&(3.9)\cr}
$$
where\footnote{Neutrino masses are not discussed here and we choose
the neutrino to be in the sneutrino mass eigenstate basis.}
the neutralino and chargino mass
eigenstates are related to the gauge eigenstates by e.g. $\wt{B} =
\sum^4_{n=1} H_{n\wt{B}}N_n,\, \wt{w}_3
= \sum^4_{n=1} H_{n\wt{w}_3} N_n,\,
\wt{w}^+ = \sum^2_{c=1} K_{c\wt{w}}\chi_c$, and
$$
\eqalignno{
W_{E_L} &= U_L^\dagger V_{E_L}, \;
W_{E_R} = U^\dagger_E V_{E_R}, \; W_{U_L}
= U^\dagger_Q V_{U_L},  \; W_{D_L} = U^\dagger_Q V_{D_L},\cr
 W_{U_R} &= U^\dagger_U
 V_{U_R},\; W_{D_R} = U^\dagger_D
V_{D_R}.\cr}
$$
\noindent There are also non-diagonal chirality-changing mass terms
$$
\eqalignno{
- {\cal{L}}^{n.d}_m &= \widetilde{e}_R^T
W^*_{E_R}(A_E + \mu \tan\beta){\boldlambda}_E
W^\dagger_{E_L} \widetilde{e}_L v_D + \widetilde{e}_R^T U^T_E
\Delta{{\boldzeta}}_E
U_L\widetilde{e}_L v_D\cr
&+ \widetilde{d}_L^T W^*_{D_L}(A_D + \mu \tan\beta)
{\boldlambda}_D W^\dagger_{D_R}
\widetilde{d}_R v_D + \widetilde{d}^T_L U^T_Q
 {\Delta{\boldzeta}}_D U_D
\widetilde{d}_R v_D\cr
&+ \widetilde{u}_L^T W^*_{U_L} (A_U + \mu\cot\beta)
{\boldlambda}_U W^\dagger_{U_R}
 \wt{u}_R v_U
+ \widetilde{u}_L^T U^T_Q {{\Delta\boldzeta}}_U
 U_U\widetilde{u}_Rv_U\cr
&+ h.c.&(3.10)\cr}
$$
The lepton flavor violating couplings are summarized in Fig.\ 2.

In the rest of this section we discuss the flavor mixing matrices
in the minimal supersymmetric standard model, minimal and general $SU(5)$
and $SO(10)$ models, with moderate or large $\tan\beta$.
The results are summarized in Table 1.

For the minimal supersymmetric standard model, the radiative corrections
 to the soft masses
only come from the Yukawa interactions of the MSSM:
$$
\eqalignno{
\Delta{\bf{m}}^2_Q &\propto
{\boldlambda}_U{\boldlambda}^\dagger_U +\kappa
{\boldlambda}_D{\boldlambda}^\dagger_D,\cr
\Delta{\bf{m}}^2_U &\propto 2{\boldlambda}^\dagger_U{\boldlambda}_U,\cr
\Delta{\bf{m}}^2_D &\propto 2{\boldlambda}^\dagger_D{\boldlambda}_D,\cr
\Delta{\bf{m}}^2_L &\propto {\boldlambda}^\dagger_E{\boldlambda}_E,\cr
\Delta{\bf{m}}^2_E &\propto 2{\boldlambda}_E{\boldlambda}^\dagger_E.
&(3.11)\cr}
$$
We have assumed a boundary condition on the scalar mass matrices
${\bf{m}}^2_{A} \propto I $ at $M_{PL}$,
and $\kappa \neq 1$ represents the possibility that the proportionality
constants are not universal.
For moderate $\tan\beta, {\lambda}_t \gg {\lambda}_b$
so that the radiative corrections are dominated by $\lambda_t$.
Thus one can neglect the ${\boldlambda}_D$ contribution and the only nontrivial
mixing is $W_{D_L}$.
For large $\tan\beta, {\lambda}_t$ and ${\lambda}_b$ are comparable, so
${\bf{m}}^2_Q$ will lie between ${\boldlambda}_U{\boldlambda}^\dagger_U$
and ${\boldlambda}_D{\boldlambda}^\dagger_D$.
Therefore both $W_{U_L}$ and $W_{D_L}$ are non-trivial.
with KM matrix and one can ignore them.

For the minimal $SU(5)$ model, there are only two Yukawa matrices,
${\boldlambda}_U = {\boldlambda}_{10}, {\boldlambda}_D = {\boldlambda}_E=
{\boldlambda}_5$, and
$$
\eqalignno{
\Delta{\bf{m}}^2_Q &\propto 3{\boldlambda}_U{\boldlambda}^\dagger_U
    + 2\kappa {\boldlambda}_D
    {\boldlambda}^\dagger_D,\cr
\Delta{\bf{m}}^2_U &\propto 3{\boldlambda}^\dagger_U{\boldlambda}_U +
    2 \kappa {\boldlambda}^\dagger_D{\boldlambda}_D,\cr
\Delta{\bf{m}}^2_D &\propto 4{\boldlambda}^\dagger_D{\boldlambda}_D,\cr
\Delta{\bf{m}}^2_L &\propto 4{\boldlambda}^\dagger_D{\boldlambda}_D,\cr
\Delta{\bf{m}}^2_E &\propto 3{\boldlambda}_U{\boldlambda}^\dagger_U +
    2\kappa{\boldlambda}_D{\boldlambda}^\dagger_D.&(3.12)\cr}
$$
For moderate $\tan\beta, {\lambda}_t \gg {\lambda}_b$,
we have non-trivial
mixings for $W_{D_L}$ and $W_{E_R}$, as found in \cite{BH,BHS}.
For large $\tan\beta$, ${\boldlambda}_D$ can not be ignored, giving
non-trivial mixings for $W_{U_L}$ and $W_{U_R}$.

For the minimal $SO(10)$ model considered in \cite{BHS,DH},
$$
\eqalignno{
\Delta{\bf{m}}^2_Q &\propto 5 {\boldlambda}_U{\boldlambda}_U^\dagger
    + 5 \kappa {\boldlambda}_D{\boldlambda}_D^\dagger,\cr
\Delta{\bf{m}}^2_U &\propto 5 {\boldlambda}_U^\dagger{\boldlambda}_U
    + 5 \kappa{\boldlambda}_D^\dagger{\boldlambda}_D,\cr
\Delta{\bf{m}}^2_D &\propto 5 {\boldlambda}_U^\dagger{\boldlambda}_U
    + 5\kappa{\boldlambda}_D^\dagger{\boldlambda}_D,\cr
\Delta{\bf{m}}^2_L &\propto 5 {\boldlambda}_U^\dagger{\boldlambda}_U
    + 5\kappa{\boldlambda}_D^\dagger{\boldlambda}_D,\cr
\Delta{\bf{m}}^2_E &\propto 5 {\boldlambda}_U{\boldlambda}_U^\dagger +
    5\kappa{\boldlambda}_D{\boldlambda}_D^\dagger.&(3.13)\cr}
$$
We have non-trivial mixings $W_{D_L}, W_{D_R}, W_{E_L}$, and $W_{E_R}$ for
moderate $\tan\beta$ and
non-trivial mixings for all $W$'s for large $\tan\beta$.

For the general $SU(5)$ or $SO(10)$ models, defined in the last
section, we get non-trivial mixings for all mixing matrices in general.
However, in $SU(5)$ models with moderate $\tan\beta$, the splittings among
${\bf{m}}^2_D$ and ${\bf{m}}^2_L$ are too small (because they
are generated by the
small ${\boldlambda}_5(A)$) to give significant flavor changing effects.

One might expect that the mixing in the $W_U$'s are smaller than those in the
$W_D$'s because of the larger hierarchy in ${\boldlambda}_U$
compared with ${\boldlambda}_D$.
However, a given $W$ is the product of a $U^\dagger$ (which diagonalizes
the scalar
mass matrix) and a $V$ (which diagonalizes the Yukawa matrix).
Even if the mixings in $V_U$'s are smaller than those in $V_D$'s because
of the larger
hierarchies in ${\boldlambda}_U$, we do not have a general argument for
the size of mixings in $U$ matrices.
This is because $U$ diagonalizes (appropriate combinations of)
known Yukawa matrices and unknown Yukawa matrices appearing
above the GUT scale, (2.5).
The mixings in $U^\dagger$ and $V$ can add up or cancel each other.
Our only general expectation is that these new Yukawa matrices have
similar hierarchical patterns as ${\boldlambda}_U$ or ${\boldlambda}_D$.
Without a specific model, one can at most say that all non-trivial $W$'s are
expected to be comparable to $V_{KM}$; the argument that the mixings in
$W_U$'s should be smaller than is $W_D$'s is not valid.

In the minimal models at moderate $\tan\beta$, the leading contributions
to flavor changing processes, such as $\mu\to e\gamma$,
involve diagrams with
a virtual scalar of the third generation.
Although such contributions are highly suppressed by mixing angles, they
dominate because they have large violations of super-GIM\cite{GIM}:
the top Yukawa coupling makes $m_{\wt{\tau}}$ very different from
$m_{\wt{e}}, m_{\wt{\mu}}$.
At large $\tan\beta$, the strange/muon Yukawa couplings get enhanced,
so the splitting between $m_{\wt{e}}$ and $m_{\wt{\mu}}$
increases, leading
to potentially competitive contributions to flavor
changing processes which do not involve the third generation.
The importance of these new diagrams can be estimated by
comparing the contributions
to $\Delta m^2_{21}$ (in a basis where gaugino vertices are diagonal) when
the super-GIM
cancellation is between scalars of the first two generations (2-1) and
third generations (3-1):
$$
{\Delta m^2_{21}(2\hbox{-}1)\over \Delta m^2_{21} (3\hbox{-}1)}
 \simeq {V_{cd}\lambda^2_2\over
V_{td}V_{ts}\lambda^2_t}
    \simeq \left\{
\matrix{ 10^{-2}, & \hbox{for}\;  \lambda_2 = \lambda_c,\cr
         \left({\tan\beta\over 60}\right)^2,
        &\hbox{for}\; \lambda_2 = \lambda_s.}\right.\eqno(3.14)
$$
We can see that for large $\tan\beta$ (or any $\tan\beta$ with small
$\lambda_s$ coming from the mixing of Higgs at $M_G$ i.e., $\lambda_s (M_G) =
{\tan\beta\over 60} \lambda_2 (M_G))$,  this could be comparable to the flavor
violating effects from the large
splitting of the third generation scalar masses.
However, for the $\mu \to e\gamma$ in $SO(10)$ models, it does not
 contribute to diagrams
which are proportional to $m_\tau$, (because it does not involve the third
 generation scalars),
the dominant contributions are still those diagrams considered
in \cite{BHS}.
For flavor changing processes which do not need chirality flipping,
such as $K-\overline{K}$ mixing, and all flavor changing processes
in $SU(5)$ models, this non-degeneracy between the first two generations is
 important.
The above discussion is summarized in Table 1.

\newpage

\centerline{\bf{\large Table 1}}

\medskip

\begin{center}
\begin{tabular}{|c|c|c|c|c|c|}
\hline
& &\multicolumn{2}{c|}{$SU(5)$} & \multicolumn{2}{c|}{$SO(10)$}\\
\hline
&MSSM&Minimal&general&minimal&general\\
\hline
\hline
$\delta m^2_3$&$\surd$&$\surd$&$\surd$&$\surd$&$\surd$\\
\hline
$\delta m^2_2$&$\bullet$&$\bullet$&$\circ$&$\bullet$&$\circ$\\
\hline
\hline
$W_{U_L}$&$\bullet$&$\bullet$&$\surd$&$\bullet$&$\surd$\\
\hline
$W_{D_L}$&$\surd$&$\surd$&$\surd$&$\surd$&$\surd$\\
\hline
$W_{U_R}$&---&$\bullet$&$\surd$&$\bullet$&$\surd$\\
\hline
$W_{D_R}$&---&---&$\surd^*$&$\surd$&$\surd$\\
\hline
$W_{E_L}$&---&---&$\surd^*$&$\surd$&$\surd$\\
\hline
$W_{E_R}$&---&$\surd$&$\surd$&$\surd$&$\surd$\\
\hline
\end{tabular}
\end{center}

\medskip
{\bf{Table 1:}} Summary table for the flavor mixing matrices:

$
\delta m^2_3$ : important effects due to some third generation
scalars not degenerate with those of first two generations.

$\delta m^2_2$ : non-negligible effects due to
 non-degeneracy of the scalars;     of the  first two generations.

$W_i$ : fermion $i$ and scalar $\wt{i}$ are rotated differently to get to mass
basis.

$\surd$ : present for any value of $\tan\beta$.

$\bullet$ : present only for large $\tan\beta$.

$\circ$ : present for large $\tan\beta$, but model dependent for
 moderate $\tan\beta$.

--: not present.

$*$ : although present, its effect for moderate $\tan\beta$
on flavor violation is small due to the small
non-degeneracy among different generation scalars.

\newpage

\noindent{\bf{IV. Phenomenology from up-type mixing}}
\medskip

As discussed in the previous section,
unlike the minimal models with moderate $\tan\beta$ studied in
\cite{BH,BHS,DH,BHS2} in generic
GUT's (for any $\tan\beta$) and even for minimal GUT's
(at large $\tan\beta$),
we expect mixing matrices in the up sector.
Having motivated an origin for non-trivial up mixing matrices
$W_{U_{L(R)}} \neq 1$,
we consider some effects they produce.
In the following we simply assume some $W_{u_{L(R)}}$
at the weak scale and consider their phenomenological consequences.
 (See however Section
V and the appendix for a discussion of
the scaling of mixing matrices from
GUT to weak scales.)
 In particular we discuss
$D - \bar{D}$ mixing, corrections to up-type quark masses,
contributions to the neutron electric dipole moment
(e.d.m.) and the possibility of different
dominant proton decay modes
than those expected from minimal models.

\noindent{\bf IVa. $D-\bar{D}$ mixing:}

To get an idea for the contribution of up-type
mixing matrices to $D-\bar{D}$ mixing, we follow
\cite{GM,NS} and employ
the mass insertion approximation.
The bounds obtained from $D-\bar{D} $ mixing on the $6\times 6$
up-squark mass matrix
$m^2_U = \pmatrix{ m^2_{U_{LL}}&m^2_{U_{LR}}\cr
m^2_{U_{RL}} & m^2_{U_{RR}} }$
(in the basis where gluino and Yukawa couplings are diagonal) are
summarized in \cite{NS}.
For average up-squark mass of $\wt{m} = 1 $ TeV, they are
$$
\sqrt{ {m^2_{U_{LL 12}}\over \wt{m}^2}
{m^2_{U_{RR 12}}\over \wt{m}^2} } \leq
.04, \eqno(4.1)
$$
$$
{m^2_{U_{LR 12}}\over \wt{m}^2} \leq  .06 . \eqno(4.2)
$$
Consider first (4.1).
In the last section we estimated that the contribution to
$m^2_{12}$ from the slight non-degeneracy between the first
two generation
scalars is generically at most comparabale to that from
 the non-degeneracy between the
first two  and
 third generation scalars.
Thus,  for our calculation,
we only consider the contribution from the
splitting between first two and third generation scalars.
Then, for $A=L, R$
$$
\abs{ {m^2_{AA 12}\over \wt{m}^2}}
= \abs{ W_{U_{A 13}} W^\dagger_{U_{A 32}}}
\abs{m^2_{\wt{t}_A}-m^2_{\wt{u}_A}\over \wt{m}^2} \leq
\abs{ W_{U_{A 13}}W^\dagger_{U_{A 32}} }.\eqno(4.3)
$$
We see that for $W$'s of the same size as the
corresponding KM matrix elements,
the LHS of (4.1) is of order $4 \times 10^{-4}$, and the bound
is easily satisfied.
Turning to (4.2), note that if ${\boldzeta}_U
= A{\boldlambda}_U,\; m^2_{U_{LR 12}} =0$.
However, we expect ${\boldzeta}_U
= A{\boldlambda}_U + \Delta{\boldzeta}_U$,
with $\Delta{\boldzeta}_U$ induced in running from $M_{PL}$ to $M_G$
having primarily  a third generation component in the gauge eigenstate basis.
If all relevant mixing matrix elements are of order the KM matrix elements,
 we expect
$\abs{ {m^2_{U_{LR 12}}\over \wt{m}^2} } = O
\left(\abs{ {Am_t\over \wt{m}^2} V_{td}V_{ts}}\right).$
Again, we see that the bound (4.2) is generically easily satisfied, and
thus we do not in general expect significant contributions to $D-\bar{D}$
mixing.

\medskip

\noindent{\bf IVb. Weak-scale corrections to up-type quark masses:}

It is well known that there are important weak-scale
radiative corrections to the
down quark mass matrix proportional to $\tan\beta$
\cite{HRS,RS,BPR,COPW,H}.
In general unified models, with non-zero $W_U$,
there are also importrant weak
scale corrections to the up quark mass matrix.

{}From the diagram in Fig.\ 3, we have a contribution to up-type
masses proportional to
$m_t$.
We find, again assuming degeneracy between the
scalars of the  first two generations,
$$
\eqalignno{
\Delta m^{ij}_u &=
{8\over 3} \left( {\alpha_s\over 4\pi}\right) m_t
\left( {A +\mu \cot\beta \over M_{\wt{g}} }\right)
W_{U_{L3i}} W_{U_{R3j}}
\left[ h (x_{t_L}, x_{t_R}) - h (x_{t_L}, x_{u_R})\right.\cr
& -\left.
h(x_{u_L}, x_{t_R}) + h(x_{u_L}, x_{u_R})\right],&(4.4)\cr}
$$
where
$$
x_i \equiv {\wt{m}^2_i\over M^2_{\wt{g}}},\;\; h(x,y) = {1\over x-y}
  \left[ {x \log x\over 1-x} - {y\log y\over 1-y}\right].\eqno(4.5)
$$
The largest fractional change in the mass occurs  for the up quark.
If $W_{U_{L(R)31}} $ is comparable to the corresponding KM
matrix element, the contribution to
${\Delta m_u\over m_u}$ is not significant.
However, if each of the $W_{U_{L(R)31}}$ are a factor 3
larger than
the corresponding KM elements we can get sizable contributions.
In Fig.\ 4, we plot ${\Delta m_u\over m_u}$ in
${m_{\tilde{u}}\over M_{\tilde{g}}}
- {m_{\tilde{t}}\over m_{\tilde{u}}}$ space, where we have assumed
$m_{\wt{u}_L} = m_{\wt{u}_R}\equiv m_{\wt{u}} ;\;
m_{\wt{t}_L} = m_{\wt{t}_R} \equiv  m_{\wt{t}}$, and we have put
$|W_{U_{L 31}}|= |W_{U_{R 31}}| = 1/30,\;
(A+ \mu \cot \beta)/m_{\wt{t}} =3$.
Any deviations from these values
can simply be multiplied in  $\Delta m_u/m_u$.
In some regions of the parameter space
it is possible to get the entire up
quark mass as a radiative effect.

\newpage

\noindent{\bf IVc. Neutron e.d.m.:}

If we attach a photon in all possible ways
to the diagram giving the contribution
to $u$-quark mass, we get a contribution to the
$u$-quark e.d.m.,
 which is proportional
to $m_t$ for any value of $\tan\beta$.
Evaluating the diagram, we find
$$
d^u = e|F|\sin\phi_u\eqno(4.6)
$$
where
$$
\eqalignno{
F=
&{8\over 3}\left({\alpha_s \over 4{\pi} }\right) m_t
{A + \mu \cot \beta\over M^3_{\wt{g}} }
W_{U_{L31}} W_{U_{L33}}^*W_{U_{R31}} W^*_{U_{R33}}\cr
&\times \bigg[
\wt{G}_2 (x_{t_{L}}, x_{t_{R}}) - \wt{G}_2(x_{t_{L}}, x_{u_{R}})
-\wt{G}_2 (x_{t_{R}}, x_{u_{L}})
+ \wt{G}(x_{u_{L}}, x_{u_{R}} )\bigg],&(4.7)
\cr}
$$
where
$$\wt{G}_2 (x, y) = {g(x) - g(y)\over x-y},\; \;  g(x) =
 {1\over 2(x-1)^3} [x^2-1-2x \log x]\eqno(4.8)
$$
and
$$
Im \left[
m_t W_{U_{L31}} W_{U_{L33}}^* W_{U_{R31}} W^*_{U_{R33}}
\right] \equiv
| m_t W_{U_{L31}}W^*_{U_{R33}}W_{U_{R31}}
W^*_{U_{R33}}|\sin \phi_u.\eqno(4.9)
$$
In general we expect
a large non-zero sin $\phi_u$.
If the combination of $W$'s appearing in the above is comparable to the
combination giving a down quark e.d.m.,
the $u$-quark contribution will dominate
over the $d$-quark contribution to the
neutron e.d.m.\ considered in \cite{DH} by a
factor ${m_t\over 4m_b\tan\beta}$,
(the factor 4 comes from the quark model result
$d_n=4/3 d_d - 1/3 d_u$).
Hence, the neutron e.d.m.\ may be competitive
with $\mu \to e\gamma$ and $d_e$ as the most
promising flavor changing signal for supersymmetric unification.

\noindent{\bf IVd. Proton decay:}

Finally we turn briefly to the relevance of up-type mixing matrices for
proton decay; in particular to the important question of the charge of the
lepton in the final state.
We know that upon integrating out the
superheavy Higgs triplets we can generate
the baryon number violating operators
${1\over 2M_H} (QQ)(QL)$ and ${1\over M_H} (EU)(DU)$
in the superpotential.
These operators must subsequently be dressed at
 the weak scale in order to
obtain four-fermion operators leading  to proton decay.
The dressing may be done with neutralinos,
charginos or gluinos where possible.
Since the dressed operator grows with
gauge couplings and vanishes for vanishing
neutralino/chargino/gluino mass,
one might naively expect gluino dressing
to be most important.
However, if the up-type mixing matrices are trivial,
gluino dressed operators can only
lead to proton decay with a neutrino
in the final state.
To see this, we examine each operator separately:
$(eu_a)(d_bu_c)\epsilon^{abc}$ (where $a,b,c$ are color indices)
must involve $u$'s from two different
generations because of the $\epsilon^{abc}$.
One of them has to be a $u$, so the other is a $c$ or a $t$.
If there is no up mixing, the up flavor does not
change in the dressing process, so the final
state would have to contain a $c$ or a $t$.
Since $m_t, m_c > m_p$, this can not happen.
Next, consider $(QQ)(QL) =
u^a_Ld^b_L(u^c_Le_L-d^c_L v_L)\epsilon_{abc}$.
By exactly the same argument as the above, the
$u_L^a d^b_L u^c_Le_L\epsilon_{abc}$
operator can not contribute to proton decay.
Thus, we see that in the absence of mixing
in the up sector, gluino dressing can
only give neutrinos in the final state.
However, the above arguments break down
if up-mixing matrices are non-trivial,
since gluino dressed diagrams give a significant contribution
to the branching ratio for charged lepton modes in proton decay.
A detailed study of flavor mixing in the up sector
\cite{A} concludes that, whether
the wino or gluino dressings are dominant,
the muon final state in proton decay
is of greatly enhanced importance.
Without the mixings, one expects
${\Gamma (p\to K^o \mu^+) \over \Gamma
(p\to K^+\bar{\nu})} \approx 10^{-3}$.
The up mixing in general models increases
this by $O(100)$ making the mode
 $p\to K^o\mu^+$
a favorable one for discovery of proton decay.

\newpage

\noindent{\bf Section V. Large $\tan\beta$: Analytic Treatment}

\medskip

The large $\tan\beta$ scenario is interesting for a number of reasons.
For moderate $\tan\beta$, the only way to understand
$m_t \gg m_b, m_\tau$ is to
have $\lambda_t \gg \lambda_b, \lambda_\tau$ at the weak scale.
This gives us little hope of attributing a
common origin for third generation
Yukawa couplings at a higher scale.
However, for large $\tan\beta \sim {{O}}\left({m_t\over m_b}\right)$,
the weak scale $\lambda_t, \lambda_b,
\lambda_\tau$ are comparable and the above
hope is restored. (In fact it is
realized in $SO(10)$ models like the ADHRS example
outlined in section VII).
For us, this is sufficient motivation
to study the large $\tan\beta$ case in
more detail. Also, this case was not
studied in \cite{BHS}. We shall see that
unexpected new features arise in the large $\tan\beta$ limit.

The largest contribution to the
$\mu \to e\gamma$ amplitude comes from
the diagram with $L-R$ scalar mass insertion (Fig.\ 5).
In the $L-R$ insertion approximation,
the amplitude for $\mu_{L(R)}$
decay is
$$
\eqalignno{
F_{L(R)} &= {\alpha\over 4\pi\cos^2\theta_W} m_\tau
W_{E_{L(R)_{32}}}
W_{E_{R(L)_{31}}}
W^*_{E_{L(R)_{33}}}
W^*_{E_{R(L)_{33}}}
    (A_E + \mu \tan\beta)\cr
&\times \left[ G_2(m^2_{\wt{\tau}_L}, m^2_{\wt{\tau}_R})
    - G_2(m^2_{\wt{e}_L}, m^2_{\wt{\tau}_R})
    -G_2 (m^2_{\wt{\tau}_L}, m^2_{\wt{e}_R})+
G_2 (m^2_{\wt{e}_L },
    m^2_{\wt{e}_R})\right],\cr}
$$
where
$$
\eqalignno{
G_2(m^2_1, m^2_2) &= {G_2(m^2_1) - G_2(m^2_2)\over m^2_1 - m^2_2},\cr
G_2 (m^2) &= \sum^4_{n =1} (H_{n\wt{B}} + \cot
\theta_W H_{n\wt{w}_3})
  g_2\left( {m^2\over M^2_n}\right).&(5.1)\cr}
$$
Note, however, that for large $\tan\beta$
the $L-R$ insertion approximation may be a bad one, since
the chirality changing mass for the third generation becomes comparable
to the chirality
conserving masses.
A correct treatment will be used for the numerical analysis
in the next section. We still expect, however, that the
amplitude to be proportional to $W_{E_{32}}W_{E_{31}}$ because of the unitarity
of
the mixing matrices: the sum of contributions from the first
 two generations is proportional
to $W_{1i}W^*_{1j} + W_{2i} W^*_{2j} =- W_{3i}W^*_{3j}$
for $i\neq j$, and the contribution from the third generation is itself
 proportional to $W_{3i}W^*_{3j}$.

Two simplifications in the dependence of the $\mu\to e\gamma$ rate on parameter
space occur for large $\tan\beta$.
First, since the dominant diagram involves the $L-R$ insertion $(A +
\mu\tan\beta)m_t$, and since $\tan\beta$ is large, the amplitude does not
depend on the weak scale parameter $A$.
Second, in the large tan$\beta$ limit, the chargino mass matrix is
$$
M_\chi = \pmatrix{ M_2 & \sqrt{2} M_W \sin \beta\cr
\sqrt{2} M_W \cos\beta &-\mu} \longrightarrow
\pmatrix{ M_2 & \sqrt{2}M_W\cr
             0&  -\mu},\eqno(5.2)
$$
and the parameters $M_2, \mu$ have a direct interpretation as the chargino
masses. (Note that this assures us that $\mu\tan\beta$ will likely always be
much bigger than $A$;
for a $\tan\beta$ of 50, the LEP lower bound on chargino mass of 45 GeV
tells us that $\mu \tan\beta > 2 $ TeV, so for $A$ to be comparable to
$\mu\tan\beta$ we must have $A > 2$ TeV.)

In considering $\mu \to e\gamma$ for large $\tan\beta$, two factors come
immediately to mind which tend to (perhaps dangerously) enhance the rate over
the case with moderate $\tan\beta$.

(i) As we have already mentioned,
the dominant contribution to $\mu \to e\gamma$
grows
with $\tan\beta$; the diagram in
Fig.\ 5 is proportional to $\tan\beta$, a
factor of 900 in the rate for $\tan\beta = 60$
compared to $\tan\beta =2$.

(ii) For large $\tan\beta,\;
\lambda_\tau$ can be $O(1)$ and we can not neglect its
contribution to the running of the slepton
mass matrix from $M_G$ to $M_S$
(soft SUSY breaking scale).
This scaling generally splits the third
generation slepton mass even further
from the first two generations, meaning a
less effective super-GIM mechanism
 and a larger amplitude for $\mu \to e\gamma$.

While both of the above effects certainly exist,
there are also two sources of
{\it{suppression}} of the amplitude for
large $\tan\beta$, which can together
largely compensate for the above factors:

(i)$'$ Large $\tan\beta$ allows $\lambda_t$ to
be smaller than for moderate
tan$\beta$.
There are two reasons for this. First,
large $\tan\beta$ allows $v_U$
to be larger and so $\lambda_t$ can be smaller to reproduce
the top mass. Secondly, $b-\tau$ unification  \cite{CEG}
is achieved with a
smaller $\lambda_\tau$ in the large $\tan\beta$ regime \cite{HRS,RS}.
Since $\lambda_t$ is smaller, a smaller non-degeneracy
between the third and first two
generations is induced in running from $M_{PL} $ to $M_G$,
suppressing the
amplitude compared to the moderate $\tan\beta$ case.

(ii)$'$ In comparing large and moderate tan$\beta$,
we must know how the
mixing matrices $W_{L,R_{3i}}$ (appearing at the
vertices of the diagrams
responsible for $\mu\to e\gamma$) compare in these two cases.
In the moderate tan$\beta$ minimal models discussed in \cite{BHS},
$W_{L,R_{3i}}$ were equal to the corresponding KM matrix elements
 $V_{KM3i}$ at
$M_G$, and this equality was approximately
maintained in running form $M_G$ to $M_S$.
As discussed in the previous sections, for more
general models  one expects that
the $W_{L(R)_{3i}}$
 at $M_G$ are equal to $V_{KM3i}$ at $M_G$ up to some
combination of Clebsches.
One might then expect (as in the minimal models)
that this relationship
continues to approximately hold at lower scales.
In fact for large $\tan\beta$ this expectation is false.
We find that often, the $W_{L(R)3i}$
{\it{decrease}} from $M_G$ to $M_S$, overcompensating
for the increased non-degeneracy between the third and first two
generation slepton masses induced by large $\lambda_\tau$
(point (ii) above).

In the following, we examine the scaling of these mixing
matrices in detail.
Consider first the lepton sector. The renormalization group
equation (RGE) for ${\boldlambda}_E$ (in the
 following
$t ={log \mu\over 16 \pi^2}$) is
$$
- {d{\boldlambda}_E\over dt} =
{\boldlambda}_E [3{\boldlambda}^\dagger_E
 {\boldlambda}_E + Tr
(3{\boldlambda}^\dagger_D{\boldlambda}_D
+ {\boldlambda}^\dagger_E{\boldlambda}_E) - 3 g^2_2 -
{9\over 5} \; g^2_1]\eqno(5.3)
$$
giving
$$
- {d\over dt}{\boldlambda}^\dagger_E {\boldlambda}_E = 6
 {\boldlambda}^\dagger_E {\boldlambda}_E + 2
{\boldlambda}^\dagger_E {\boldlambda}_E Tr
(3{\boldlambda}^\dagger_D {\boldlambda}_D +
{\boldlambda}^\dagger_E{\boldlambda}_E) - (6g^2_2+{18\over5} \; g^2_1)
{\boldlambda}^\dagger_E {\boldlambda}_E \eqno(5.4)
$$
$$ - {d\over dt} {\boldlambda}_E {\boldlambda}_E^\dagger =
 6 {\boldlambda}_E{\boldlambda}_E^\dagger + 2
{\boldlambda}_E{\boldlambda}^\dagger_E Tr (3{\boldlambda}^\dagger_D
{\boldlambda}_D +
{\boldlambda}^\dagger_E{\boldlambda}_E) - (6g^2_2 + {18\over 5}\; g^2_1)
{\boldlambda}_E {\boldlambda}_E^\dagger . \eqno(5.5)
$$
These in turn imply that the basis in which ${\boldlambda}^\dagger_{{E}}
{\boldlambda}_{{E}}$
is diagonal, and the (in general different) basis where
${\boldlambda}_{{E}}{\boldlambda}_{{E}}^\dagger$
 is diagonal, do not change with scale.
Consider now the evolution of the left handed slepton mass matrix
${\bf{m}}^2_L$.
 The RGE for ${\bf{m}}^2_L$ is
$$
-{d\over dt} {\bf{m}}^2_L =
({\bf{m}}^2_L + 2m^2_{H_d}){\boldlambda}^\dagger_E
{\boldlambda}_E + 2
{\boldlambda}^\dagger_E {\bf{m}}^2_E
{\boldlambda}_E + {\boldlambda}^\dagger_E
{\boldlambda}_E
 {\bf{m}}^2_L + 2 {\boldzeta}_E^\dagger
{\boldzeta}_E + \; \hbox{Gaugino
terms} .\eqno(5.6)
$$
In the basis where ${\boldlambda}^\dagger_{{E}}
{\boldlambda}_{{E}}$ is diagonal,
 keeping only the $\lambda_\tau$ contribution, the $3i$
entry $(i\neq 3)$ becomes:
$$
- {d\over dt} m^2_{L3i} = \lambda^2_{\tau} m^2_{L3i} + 2
 ({\boldzeta}^\dagger_E {\boldzeta}_E)_{3i} .\eqno(5.7)
$$
In this basis, we have ${\bf{m}}^2_L
= W^\dagger_L\overline{\bf{m}}_L^2W_L$.
(Here and in the remainder of this section, we abbreviate
$W_{E_{L(R)}}\to W_{L(R)}$).
Assuming degeneracy between scalars of  the first two generations,
$
 m^2_{L_{3i}} = W_{L 3i}W_{L33}^\dagger
(m^2_{\tau_{L}}- m^2_{e_{L}}) \equiv W_{L3i}W^\dagger_{L 33} \Delta m^2_L.
$
Then (5.7) becomes
$$
- {d\over dt} (W_{L3i} W_{L33}^\dagger\Delta m^2_L)
= \lambda^2_{\tau} (W_{L3i}W^\dagger_{L 33}
\Delta m_L^2) + 2({\boldzeta}^\dagger_E {\boldzeta}_E)_{3i} .\eqno(5.8)
$$
For now, we ignore the $({\boldzeta}^\dagger_E{\boldzeta}_E)_{3i}$
term in (5.8), yielding the solution:
$$
(W_{L3i}W_{L33}^\dagger \Delta m^2_L)(M_S) = e^{-I_\tau} (W_{L3i}
W_{L33}^\dagger \Delta m^2) (M_G),
\eqno(5.9)
$$
where
$$
I_i \equiv \int_0^{\log {M_G\over M_S}}
{dt\over 16\pi^2} \lambda^2_i (t).
\eqno(5.10)
$$
Thus,
$$
W_{L 3i} W^\dagger_{L33}(M_S) =
e^{-I_\tau} {\Delta m^2_R(M_G)\over \Delta
m^2_R (M_S)}
W^\dagger_{L3i}W_{L_{33}} (M_G).\eqno(5.11)
$$
Similarly, we find
$$
W_{R 3i} W^\dagger_{R33} (M_S) = e^{-2I_\tau}
 {\Delta m^2_R (M_G)\over \Delta m^2_R (M_S)}
W_{R 3i}W^\dagger_{R_{33}} (M_G).\eqno(5.12)
$$
Note that, generically the quantities
${\Delta m^2_{L(R)}(M_G)\over \Delta
m^2_{L(R)} (M_S)}$ are smaller than one, since the third
generation mass
gets split even further from the first
two generations in running from $M_G$ to
$M_S$.
Thus, we find that the $W_{L(R)_{3i}}$
 get smaller in magnitude as we scale from
$M_G$ to $M_S$, in contrast with the KM matrix elements $V_{KM3i}$, which scale
as
$$
V_{KM3i} (M_S) = e^{(I_t+I_b)} V_{KM 3i} (M_G).\eqno(5.13)
$$
Suppose that at $M_G$ the
$W_{L(R)}$ are related to $V_{KM}$
though some combination of Clebsches
determined by the physics above the
GUT scale.
$$
W^\dagger_{L(R)33}W_{L(R)3i}
 (M_G) = z_{i{L(R)}} V_{KM 3i} (M_G) .\eqno(5.14)
$$
This relationship is not maintained at lower scales; instead we have:
$$
\hspace{-.2in}
W^\dagger_{L33}W_{L 3i} (M_S) = {\Delta m^2_L(M_G)\over \Delta m^2_L (M_S)}
e^{-(I_\tau+I_t+I_b)}
    z_{i_{L}}V_{KM 3i}(M_S), \eqno(5.15)
$$
$$
W^\dagger_{R33}W_{R 3i} (M_S) = {\Delta m^2_R(M_G)\over \Delta m^2_R (M_S)}
 e^{-(2I_\tau + I_t+I_b)} z_{i_{R}}V_{KM 3i}(M_S).\eqno(5.16)
$$
The dominant contribution to the
$\mu \to e\gamma$
rate is proportional to \newline
$| W^\dagger_{L33} W_{L{32}}W^\dagger_{R33}  W_{R{31}} (M_S)|^2 +
 | W^\dagger_{L33} W_{L{31}}W^\dagger_{R33} W_{R{32}}(M_S)|^2$,
 giving
$$
\eqalignno{
Br( \mu\to e\gamma) &=
\left[
    {\Delta m^2_L (M_G)\over \Delta m^2_L (M_S)}
    {\Delta m^2_R (M_G)\over \Delta m^2_R (M_S)}
\right]^2
    e^{-(6I_\tau + 4I_t + 4I_b)}
\times (|z_{2_L}z_{1_R}|^2 + |z_{1_L}z_{2_R}|^2) \cr
&\times
    Br(\mu\to e\gamma, W_{L(R)3i}
    W^\dagger_{L(R)33}(M_S)\to V_{KM3i}(M_S))\cr
&\hspace{-.7in}
\equiv \epsilon (|z_{2_L} z_{1_R}|^2 + | z_{1_L} z_{2_R}|^2)
\times Br(\mu\to e\gamma, W^\dagger_{L(R)33}
    W_{L(R)33}(M_S)\to V_{KM3i}(M_S)).\cr
&&(5.17)
\cr}
$$
This $\epsilon$ represents a possibly significant suppression
of the rate for large $\tan\beta$.

At this point, the reader may object: it is true that
the $W_{L(R)3i}$ decrease from $M_G$ to $M_S$, but as already mentioned,
the non-degeneracy between the third and first two generating is increasing.
Which effect wins?
We argue that in general there is a net suppression.
This is easiest to see if in computing the $\mu \to e\gamma$ amplitude,
we use the mass insertion approximation rather than mixing
matrices  at the the vertices (Fig.\ 6).
Although this may be a poor approximation,
it serves to illustrate our point.
(Of course no such approximation is made in our numerical work.)
{}From the diagram it is clear that the amplitude is proportional to
 $m^2_{L32} m^2_{R31} (M_S)$.
{}From (5.7), we see that the rate scales as
$$
\left( m^2_{L32}m^2_{R31}\right)^2 (M_S) = e^{-(6I_\tau +4I_t+4I_b)}
\left(m^2_{L_{32}}m^2_{R_{31}}\right)
(M_G),\eqno(5.18)
$$
a net suppression.
In the mass insertion approximation,  then, the terms
${\Delta m^2(M_G)\over \Delta m^2(M_S)}$
in (5.17) serve to exactly compensate
for the increased non-degeneracy between $m^2_{\wt{e}_L}$
 and $m^2_{\wt{\tau}_L}$; what remains is
still a suppression.
This, together with (i)$'$ invalidates the naive
expectation that the theory
is ruled out in most regions of parameter space due
to the enhancing factors (i)
and (ii), (although there are still stringent
constraints on the parameter space).

The above analysis suggests that individual
lepton number conservation
is an infrared fixed point of the MSSM (whereas individual quark
number conservation is
an ultraviolet fixed point).
A more complete analysis of scaling for the lepton sector
and a discussion of scaling in the quark sector
is presented in the appendix.

\newpage

\noindent{\bf Section VI. Large $\tan\beta$: Numerical Results}

\medskip

The amplitude for $\mu \to e\gamma$ depends on the $6\times 6$ slepton mass
matrix ${\bf{M}}^2$.
In the basis where ${\bf{m}}^2_L,{\bf{ m}}^2_E$ are diagonal, we have
$$
{\bf{M}}^2 = \pmatrix{
{\bf{\overline{m}}}^2_{{E}_{L}} + {\bf{D}}_L & {\bf{k}}\cr
{\bf{k}}^\dagger &{\bf{\overline{m}}}^2_{{E}_{R}} + {\bf{D}}_R}\eqno(6.1)
$$
where in the large $\tan\beta$ limit, $D_i = - (T_{3i} - Q_i
 \sin^2\theta_W)M^2_Z$
is the $D$-term contribution,
and $k_{ij} = \mu m_\tau \tan\beta W_{L 3i}W_{R 3j}$.
The amplitude from Fig.\ 1 for $\mu_L$ decay is
$$
F_L = {\alpha\over 4\pi\cos^2\theta_W}
W^\dagger_{L i2} G_2({\bf{M}}^2)_{LR ij}
    W^\dagger_{Rj1}
$$
where
$$
G_2 ({\bf{M}}^2) \equiv \pmatrix{G_2(M^2)_{LL} & G_2(M^2)_{LR}\cr
                  G_2(M^2)_{RL} & G_2(M^2)_{RR} }, \eqno(6.3)
$$
In \cite{BHS}, ${\bf{M}}^2$ was
approximately diagonalized by the $\mu m_\tau
\tan\beta$ insertion approximation,
and $G_2 ({\bf{M}}^2)$ was calculated using
this approximate diagonalization.
Since here $\tan\beta$ is large,
we wish to avoid making such an approximation,
and numerically diagonalize the full 6$\times 6\; {\bf{M}}^2$.

Faced with a rather large parameter space,
we must decide which parameters to use in
our numerical work.
We have firstly decided to do our analysis only for large
$\tan\beta$, since the moderate
$\tan\beta$ scenario has been covered in \cite{BHS}.
Secondly, we choose to present our results
in a different way than in \cite{BHS},
where the rates for $\mu \to e\gamma$ were
plotted against a combination of
Planck scale
and weak scale parameters.
In our work, we compute $\mu \to e\gamma$
entirely in terms of weak scale parameters.
In particular, we assume that the necessary
condition for a significant
 $\mu\to e\gamma$ rate exist at the weak
scale, namely non-trivial mixing
matrix $W_{L,R_{3i}}$
and non-degeneracy between third and first two generation
slepton masses.
In the previous sections, we have shown a possible way
in which these
ingredients may be produced.
Our plots for $\mu \to e \gamma$ rates are
made against low energy parameters,
and we separately plot the regions
in low energy parameter space predicted
by our particular scenario for generating
$\mu\to e \gamma$.
This way, our plots are in terms of
experimentally accessible quantities
and can be thought of as constraining
the parameter space of the effective 3-2-1
softly broken supersymmetric theory resulting from the spontaneous
breakdown of a GUT.
(We use the GUT to relate weak scale gaugino masses.)
Our low energy plots have no dependence on the physics above the GUT scale,
all the model dependence comes into the
predictions for low energy parameters the GUT
makes.
If the predicted region of low energy parameters
corresponds to a $\mu \to e\gamma$
rate exceeding experimental bounds, the theory is ruled out.

There is a more practical reason for working
directly with low-energy parameters
specific to large $\tan\beta$: the well known
difficulty in achieving electroweak symmetry breaking in this regime.
Working with high energy parameters,
and imposing universal scalar masses necessitates a fine-tune to achieve
$SU(2) \times U(1)$ breaking. However, we
have nowhere in our analysis made the
assumption of universal scalar masses,
hence the Higgs masses and squark/slepton
masses are independent in our analysis,
and therefore the $\mu$ parameter is
not tightly constrained by squark/slepton masses.
Working with weak scale parameters allows us to assume that the desired
breaking has occurred without having to know the details of the breaking.

With the aforementioned assumption about the existence of a GUT, and
assuming degeneracy between the first two generations, the rate for
$\mu \to e\gamma$
depends on the weak scale parameters
$\mu, \tan\beta, M_2, m^2_{\wt{e}_L}, m^2_{\wt{\tau}_L},
m^2_{\wt{e}_R}, m^2_{\wt{\tau}_R}, $
$W_{L 3i}, W_{R 3i}$.
We know that the amplitude depends on
$W_{L(R)3i} $ simply through the product
$W_{L 3i} W_{R 3j}$, so for normalization in our plots we put
$W_{L(R)3i} = V_{KM 3i}$.
Any deviation from this can be simply multiplied into the rate.
We also fix $\tan\beta = 60$,
and put $m_{\wt{\tau}_{L(R)}} = m_{\wt{e}_{L(R)}} -
\Delta_{L(R)}$.
Next, we use some high energy bias to relate $m_{\wt{e}_L}$
and $m_{\wt{e}_R}$:
we assume that their difference is proportional to $M_2$
( as would be the case if they
started out degenerate and were split only through
different gauge interactions),
 so we put $m_{\wt{e}_L}= m_{\wt{e}_R} - rM_2$.
In all specific models we have looked at, $r$ is small
(less than about .2).
We find that, as long as  $r$ is small, the rate has little dependence on
 its exact value, so we put
$r=0, m_{\wt{e}_L} = m_{\wt{e}_R} \equiv \overline{m}_{\wt{e}}$.
We also found that as long as
${\Delta_L\over \Delta_R}$ is close to 1, there is
little dependence on its actual value either, so we put
$\Delta_L=\Delta_R\equiv \Delta$.

Now, the $\mu \to e\gamma$ rate depends
only on $\mu, M_2, \overline{m}_{\wt{e}}$ and
$\Delta$,
and we have the large $\tan\beta$ interpretation of $\mu$ and $M_2$
as chargino masses.
Fixing $\overline{m}_{\wt{e}}=300$ GeV, we make contour plots
of $Br(\mu \to e\gamma)$.
The rate scales roughly as ${\overline{m}_{\wt{e}}}^{-4}$ and $\mu^2$
for scalar masses heavy compared with gaugino masses.
In Fig.\ 7, we fix $\mu$ and plot in $M_2-\Delta$ space.
In Fig.\ 8, we fix $\Delta$ and plot in $\mu - M_2$ space.
In Fig.\ 9, we plot the values of
$\Delta$ predicted by the GUT against $M_2$, for various
values of $\lambda_t (M_G)$
and $A_e(M_S)$ and for two values of $b_5$, the gauge beta function
 coefficient above the
GUT scale.
In Fig.\ 10, we plot the suppression
factor $\epsilon$ for the same parameter
set as in Fig.\ 9.
We see that, over a significant region in parameter space,
$\epsilon$ is small, between 0.2 and 0.01.

It is clear from Fig.\ 7 that, with no suppression,
a typical value for $\Delta$ of
0.3 ($\times$300GeV)
would give rise to rates above the current bound of
$Br(\mu \to e\gamma) < 4.9 \times 10^{-11}$\cite{B}.
However, from Fig.\ 10, the suppression from $\epsilon$
is seen to be typically 20, allowing $ \Delta$'s
of up to 0.45 ($\times$300GeV).
 We see that $\epsilon $ is crucial in giving the GUT more
breathing room, as $\Delta$'s of less  than 0.45 are more common.
{}From Fig.\ 8 it is also clear that
regions of small $\mu$ and $M_2$ (that is,
light chargino masses) are preferred.
Smaller $\mu$ is preferred because it decreases the L-R mass
$\mu m_\tau \tan\beta$,
 small $M_2$ is preferred because
in the limit that the neutralino mass
tends to zero,
the diagrams Fig.\ 5 vanish.
We also note that smaller $\mu, M_2$
are preferred for electroweak symmetry
breaking\cite{HRS,RS}.

If $\mu$ and $M_2$ are both small, the
lightest supersymmetric particle (LSP)
 can be quite light,
(but where it has significant higgsino component,
it must be heavier than 45 GeV
in order to be consistent with the precise
measurement of the $Z$ width), and it
annihilates (primarily through its higgsino components) through a $Z$
into fermion antifermion pairs much like a heavy neutrino.
The contribution of the LSP to energy density of the universe
 $\Omega h^2$ then just depends on its mass, and the size of
its higgsino components, both of which only depend on
$\mu$ and $M_2$ in the large $\tan\beta$
limit.
In Fig.\ 11, we make a plot of $\Omega h^2$ in $\mu-M_2$
space. We see that it is possible to get $\Omega
 \sim O(1)$ in some region of the
parameter space.

\newpage

\noindent{\bf{VII. The Example of ADHRS Models}}

\medskip

In this section, we study the ADHRS models \cite{ADHRS} which are known to
give realistic fermion masses and mixing patterns.
These models are specific enough for us to do calculations and
make some real predictions.
Although not necessarily correct, they are good representatives of
general GUT models.
 We believe that by studying them,
one can see in detail the general features of generic realistic GUT models and
the differences between them and the minimal $SU(5)$ or $SO(10)$ models.

As mentioned in Sec.\ II, in ADHRS models, the three families of
quarks and leptons lie in
three 16 dimensional representations of $SO(10)$, and the two low
energy Higgs doublets lie in a single 10 dimensional representation.
Only the third generation Yukawa couplings come from a renormalizable
interaction
$$
\lambda_{33} 16_3 16_3 10 .\eqno(7.1)
$$
All other small Yukawa couplings come from nonrenormalizable interactions
after integrating out the heavy fields.
These interactions can be written in general as
$$
16_i \lambda_{ij} (A_a) 16_j 10.\eqno(7.2)
$$
The $A_{a}$'s are fields in the adjoint representation of $SO(10)$
 and their
vevs break $SO(10)$ down to the Standard Model gauge group.
Therefore, these Yukawa couplings can take different values for
fermions of the same generation
with different quantum numbers under $SU(3)\times SU(2) \times U(1)$
and a realistic fermion mass pattern and nontrivial KM matrix can be generated.
In ADHRS models, the minimal number (four) of operators is assumed to
generate the up,
down-type quark and charged lepton Yukawa coupling matrices
${\boldlambda}_U, {\boldlambda}_D$
and ${\boldlambda}_E$, and they take the form at $M_G$
$$
{\boldlambda}_U = \pmatrix{
0    &z_uC &0\cr
z'_uC&y_u E & x_uB\cr
0    & x'_uB& A},
 {\boldlambda}_D =
 \pmatrix{
0    &z_dC   &0\cr
z'_dC & y_dE  & x_dB\cr
0    & x'_dB & A},
{\boldlambda}_E=\pmatrix{
0    & z_eC  & 0\cr
z'_eC& y_eE  & x_eB\cr
0    & x'_eB & A},
\eqno(7.3)
$$
where the $x, y, z$'s are Clebsch factors arising from the VEV's of the
adjoint Higgs fields $A_a$.
This form is known to give the successful relations $V_{ub}/V_{cb}
 = \sqrt{m_u/m_c}$
and $V_{td}/V_{ts} = \sqrt{ m_d/m_s}$ \cite{HR} so it is well motivated.
Strictly speaking, the interaction (7.2) become the usual Yukawa
form only after the adjoints $A_a$ take their VEV's
at the GUT scale.
However, as we explained in Sec.\ II, they can be treated
 as the usual Yukawa interactions up to the
ultraheavy scale (which
we will assume to be $M_{PL}$) where the ultraheavy fields are integrated out
if the wavefunction renormalizations of $A_a$'s are ignored.
In the one-loop approximation which we use later
in calculating radiative corrections from $M_{PL}$ to $M_G$, they give
the same results, because the wavefunction
renormalizations of the adjoints $A_a$ only contribute at  the
two-loop order.
This makes our analysis much easier.
Above the GUT scale, in addition to the Yukawa interactions (2.4) which give
the fermion masses, we have the  interactions (2.5) as well.
Each Yukawa matrix has different Clebsch factors $x, y, z$
associated with its elements.
All the Yukawa matrices have the ADHRS form
$$
\lambda_I = \pmatrix{ 0& z_I C & 0\cr
z'_I C & y_IE & x_IB\cr
0 & x'_I B & A },\; I = qq, eu, ud, q\ell, nd, n\ell .\eqno(7.4)
$$
If each entry of the Yukawa matrices is generated dominantly by a single
operator,
like in the ADHRS models, then the phases of the same entries of all Yukawa
matrices are identical.
One can remove all but the $\lambda_{22}$ phases by rephasing the operators.
After phase redefinition only $E$ is complex and is responsible for
 CP violation.
In order to generate the realistic fermion mass and mixing pattern,
one expects the following hierarchies,
$$
\eqalignno{
{B\over A} &\sim V_{cb}         \hskip .14in\sim \epsilon^2, \cr
{E\over A} &\sim {m_s\over m_b} \hskip 8pt\sim \epsilon^2, \cr
{C\over E} &\sim \sin \theta_c  \sim \epsilon,
\;\;\; \hbox{where } \; \epsilon  \sim 0.2.&(7.5)\cr}
$$
The hierarchical Yukawa matrices can be diagonalized approximately
 \cite{HR},
the unitary rotation matrices which diagonalize them at
the GUT
scale can be approximately written as
$$
{\boldlambda} =\pmatrix{ 0& zC & 0\cr
z'C & y\abs{E}e^{\widetilde{\phi}}&xB\cr
0 & x'B&A } = V^*_F \bar{\boldlambda} V^\dagger_B,\eqno(7.6)
$$
$$
V_F \simeq \pmatrix{ e^{i\phi} & S_{F_1} e^{i\phi} & 0\cr
                -S_{F_1} & 1 & S_{F_2}\cr
S_{F_1}S_{F_2} & -S_{F_2} & 1\cr},\eqno(7.7)
$$
$$
V_B \simeq \pmatrix{
1                           & S_{B_1}            & 0\cr
-S_{B_1}  e^{-i\phi} & e^{-i\phi}& S_{B_2}   \cr
S_{B_1}  S_{B_2} e^{-i\phi} & -S_{B_2}e^{-i\phi} & 1},\eqno(7.8)
$$
where
$$
\eqalignno{
S_{F_2}
        &= {xB \over A},\; S_{B_2} = {x'B \over A},\cr
S_{F_1}
        &= {zC \over E'},\; S_{B_1} = { z'C \over E'},\cr
E'      &= \abs{ y E -  S_{F2} S_{B2} A  }
        =
           \abs{ y |E| e^{i\wt{\phi} } - {x'x B^2\over A} },\cr
\wt{\phi} &= arg (E),\;
\phi      = arg (y E - {x'x B^2\over A} ).
}
$$
The soft SUSY-breaking scalar masses for the three low energy generations
and trilinear $A$ terms are
assumed to be universal\footnote{If the nonrenormalizable operators
already appear in the superpotential of the underlying
supergravity theory, the $A$ terms will be different for different dimensional
operators, and will induce unacceptably large $\mu \to e\gamma$
rate because the triscalar interactions  and the Yukawa interactions can not
be diagonalized in the same basis for the first two generations.
In theories where the nonremormalizable operators come from
integrating out heavy fields at $M_{PL}$ and all the  relevant
interactions have the same
$A$ term, the resulting nonrenormalizable operators will also have
the same $A$ term.}
 at Planck scale $M_{PL}$ as in
\cite{BHS}.
Beneath $M_{PL}$,
the radiative corrections from the Yukawa couplings destroy the
universalities and render the mixing matrices non-trivial.
In the one-loop approximation, the radiative corrections to the
soft SUSY-breaking parameters at $M_{G}$ are simply related
to the Yukawa coupling matrices and therefore the relations between
general mixing matrix elements and KM matrix elements are also simple.
This allows us to see the similar hierarchies in the general mixing
matrices and the KM matrix very clearly.
Although the one-loop approximation may not be a good approximation for
 quantities involving third generation Yukawa couplings,
we will be satisfied with it since it simplifies things a lot and the
uncertainties in other quantities such as Clebsch factors are
probably much bigger than the errors made in the one-loop approximation.
The RG equations, for ${\bf{m}}^2_E$ as an example, from $M_{PL}$ to
$M_G$ are
$$
\eqalignno{
{d\over dt} {\bf{m}}^2_E &= {1\over 16\pi^2}
\bigg[2(2{\boldlambda}_E{\bf{m}}^2_L
{\boldlambda}^\dagger_E + 2 {\boldlambda}_E m^2_{H_D}
{\boldlambda}^\dagger_E
    + {\bf{m}}^2_E
 {\boldlambda}_E {\boldlambda}^\dagger_E +
{\boldlambda}_E{\boldlambda}_E^\dagger {\bf{m}}^2_E + 2
   {\boldzeta}_E{\boldzeta}^\dagger_E)\cr
&+ 3(2 {\boldlambda}_{eu} {\bf{m}}^2_U {\boldlambda}^\dagger_{eu}
+ 2 {\boldlambda}_{eu} m^2_{H_{U_3}} {\boldlambda}^\dagger_{eu}
+ {\bf{m}}^2_E {\boldlambda}_{eu}
{\boldlambda}^\dagger_{eu} + {\boldlambda}_{eu} {\boldlambda}^\dagger_{eu}
 {\bf{m}}^2_E + 2{\boldzeta}_{eu}
{\boldzeta}^\dagger_{eu})\cr
&- \hbox{gaugino mass contribution}\bigg].&(7.9)\cr}
$$
In the one-loop approximation, the gaugino mass contributions are diagonal
and the same for all three generations, so they can be absorbed into
the common scalar masses and do not affect the diagonalization.
The corrections to scalar masses at $M_G$ have the following leading
flavor dependence
$$
\eqalignno{
\Delta {\bf{m}}^2_E &\propto 2 {\boldlambda}_E{\boldlambda}_E^\dagger + 3
    {\boldlambda}_{eu} {\boldlambda}^\dagger_{eu}\cr
&= 5 \pmatrix{ \overline{z^2_e} C^2 & \overline{z_e y_e}CE^* &
            \overline{ z_ex_e}CB\cr
               \overline{z_ey_e} CE & \overline{z'^{2}_e}C^2 +
    \overline{y^2_e} \abs{E}^2 + \overline{x^2_e} B^2 &
                \overline{ y_ex'_e}EB + \bar{x}_e BA\cr
               \overline{z_ex'_e}  CB & \overline{y_ex'_e} E^*B +
        \bar{x}_e BA &
               \overline{x'^2_e} B^2 + A^2},&(7.10)\cr}
$$
 where the overline represents the weighted average of the Clebsch factors,
$\overline{z^2_e} = {1\over 5} (2z^2_e + 3z^2_{eu})$ and so on.
Because $\Delta {\bf{m}}^2_E$ is hierarchical, assuming no big $x, y$
Clebsches (ADHRS models have some
big $z$ Clebsches), the rotation matrix which diagonalizes it can be
given approximately as
$$
U_E(M_G) \simeq
\pmatrix{
1 & \bar{S}_{E_1} & \bar{S}_{E3}\cr
                          -\bar{S}_{E_1} e^{-i\widetilde{\phi}}
        & e^{-i\wt{\phi}} & \bar{S}_{E3}\cr
                          \bar{S}_{E_1}\bar{S}_{E_2}e^{-i\wt{\phi}} -
                          \bar{S}_{E3} & -\bar{S}_{E2} e^{-i\wt{\phi}} & 1},
\eqno(7.11)
$$
where
$$
\eqalignno{
\bar{S}_{E_2} &= {\bar{x_e}B\over A},\; \bar{S}_{E_3}
    = { \overline{z_ex'_e}CB\over A^2}\cr
\bar{S}_{E_1} &= { \overline{z_ey_e} C\abs{E}\over
                   \overline{z'^{2}_e} C^2  + \overline{y^2_e} \abs{E}^2 +
                  (\overline{x^2_e} - \overline{x}^2_e)B^2 },\;
                  e^{-i\widetilde{\phi}} = {E\over \abs{E} }.
\cr}
$$
Similarly, for other scalar masses the leading flavor dependent corrections
at $M_G$ are
$$
\eqalignno{
\Delta {\bf{m}}^2_L &\propto {\boldlambda}^\dagger_E{\boldlambda}_E
    + 3{\boldlambda}^\dagger_{q\ell}{\boldlambda}_{q\ell} +
    {\boldlambda}^\dagger_{n\ell}{\boldlambda}_{n\ell},\cr
\Delta{\bf{m}}^2_Q &\propto {\boldlambda}_U{\boldlambda}_U^\dagger
    + {\boldlambda}_D{\boldlambda}^\dagger_D
    + 2{\boldlambda}_{qq}{\boldlambda}^\dagger_{qq}
    + {\boldlambda}_{q\ell}{\boldlambda}^\dagger_{q\ell},\cr
\Delta {\bf{m}}^2_U &\propto 2 {\boldlambda}^\dagger_U{\boldlambda}_U
    + {\boldlambda}^\dagger_{eu}{\boldlambda}_{eu}
    + 2 {\boldlambda}_{ud}{\boldlambda}^\dagger_{ud},\cr
\Delta{\bf{m}}^2_D &\propto 2{\boldlambda}^\dagger_D{\boldlambda}_D
    + {\boldlambda}^\dagger_{nd}{\boldlambda}_{nd} +
    2{\boldlambda}^\dagger_{ud}{\boldlambda}_{ud},&(7.12)
\cr}
$$
and the rotation matrices which diagonalize them are given by
expressions similar to (7.11) with Clebsches
replaced by the appropriate ones.
Then, the mixing matrices appearing at the lepton-slepton-gaugino
vertices are given by

$$
\eqalignno{
&W_{E_L}(M_G) = U^\dagger_L V_{E_L}
\cr
&\simeq
\pmatrix{
1 & S_{E_{L1}} - {S}_{L1} e^{i(\wt{\phi} - \phi_e)} &
        \bar{S}_{L_1} (\bar{S}_{L_2} -S_{E_{L_2}}) e^{i\wt{\phi}}
        - \bar{S}_{L_3}\cr
        (\bar{S}_{L_1}- \bar{S}_{E_{L_1}}) e^{-i(\wt{\phi}-\phi_e)} &
        e^{i(\wt{\phi}-\phi_e)} & -(\bar{S}_{L_2} - S_{E_{L_2}})
        e^{i\wt{\phi} }\cr
-S_{E_{L_1}}(\bar{S}_{L_2} - S_{E_{L_2}}) e^{-i{\phi}_e} + \bar{S}_{L_3} &
        (\bar{S}_{L_2}-S_{E_{L_2}}) e^{-i\phi_e} & 1
},\cr
&&(7.13)\cr}
$$
$$
\eqalignno{
&W_{E_L}(M_G) = U^\dagger_E V_{E_R}
\cr
&\simeq
\pmatrix{
e^{i\phi_e}
& S_{E_{R1}}e^{i{{\phi}}} e-\bar{S}_{E_{1}}e^{i\wt{\phi}}&
        \bar{S}_{E_1}(\bar{S}_{E_2} - S_{E_{R2}})
        e^{i\wt{\phi}} - \bar{S}_{E_3}\cr
        \bar{S}_{E_1}e^{i\phi_e}-S_{E_{R1}}e^{i\wt{\phi}} &
        e^{i\wt{\phi}} & - (\bar{S}_{E_2}- S_{E_{R_2}})e^{-i{\phi}}\cr
      -{S}_{E_{R1}}(\bar{S}_{E_2} - S_{E_{R2}})+ \bar{S}_{E_3} e^{i{\phi_e}} &
        \bar{S}_{E_2} - S_{E_{R2}} &1},\cr
&&(7.14)\cr}
$$
\noindent where
$$
\eqalignno{
S_{E_{L_1}} &=
    {z'_e C\over E'_e},\;  S_{E_{R_1}}
    = {z_e C\over E'_e},\;
    E'_e =  |y_e E - {x'_e x_e B^2\over A}|,\cr
S_{E_{L_2}} &= {x'_e B\over A},\;
     S_{E_{R_2}} = {x_e B\over A},\;
    \phi_e =
    arg \left( y_e\abs{E} e^{i\wt{\phi}} -
{x'_e x_e B^2\over A}\right)\cr
& {\hskip 1.75in}
    \simeq \wt{\phi}\;\; (\hbox{if}\; y_e \sim x_e, x'_e),\cr
\bar{S}_{E_1} &= { \overline{z_ey_e} C |E|\over \overline{z'^2_e} C^2
+ \overline{y^2_e} \abs{E}^2
+ (\overline{x^2_e} - \overline{x}^2_e) B^2},
    \; \bar{S}_{E_2}={\bar{x}_eB\over A}, \;
 \bar{S}_{E_3} =
{\overline{z_e x'_e} C B\over A^2},\cr
\bar{S}_{L_1} &= { \wh{z'_e y_e} C|E|\over \wh{z^2_e} C^2 +
\wh{y^2_e}
    |E|^2+ (\wh{x^2_e} - \wh{x}^2_e)B^2 },\;
    \bar{S}_{L_2} =
    {\wh{x'_e} B\over A},\;
    \bar{S}_{L_3} = { \wh{z'_e x_e} CB\over A^2}, \cr
\hbox{and}&\cr
\bar{x}_e & = {1\over 5} (2x_e + 3 x_{eu}),\;\; \wh{x'_e}= {1\over 5}
(x'_e + 3 x'_{g\ell} + x'_{n\ell}) \;\; \hbox{etc.}
\cr}
$$
Note that
$$
\bar{S}_{L_3} \sim \ {\hbox{Clebsch}}\ \times
 {CB\over A^2},\;  S_{E_{L_1}} (\bar{S}_{L_2}-\bar{S}_{E_{L_2}})
= {\hbox{Clebsch}} \times {CB\over EA}.\eqno(7.15)
$$
If there is no very big or small Clebsch involved and
no accidental cancellation,
$\bar{S}_{L_3}, \bar{S}_{E_3}$ can be neglected in $W$'s.

Compared with $V_{KM}$,
$$
\eqalignno{
&V_{KM} (M_G) = V^\dagger_{U_L} V_{D_L}
\cr
&\simeq
\pmatrix{
1        & S_{D_{L_1}}-S_{U_{L_1}} e^{-i(\phi_d-\phi_u)} & -S_{U_{L_1}}
        (S_{D_{L_2}}-S_{U_{L_2}}) e^{i\phi_u}\cr
S_{U_{L_1}} - S_{D_{L_1}} e^{-i(\phi_d-\phi_u)} & e^{-i(\phi_d -\phi_u)}
    & (S_{D_{L_2}} - S_{U_{L_2}}) e^{i\phi_u}\cr
S_{D_{L_1}} (S_{D_{L_1}} -S_{U_{L_2}}) e^{-i\phi_d} & -(S_{D_{L_2}} -
 S_{U_{L_2}}) e^{-i\phi_d}    & 1 },\cr
&&(7.16)\cr}
$$
where
$$
\eqalignno{
S_{U_{L1}} &= {z_u C\over E'_u},\;
E'_u = \abs{ y_u E-{x_u x'_uB^2\over A} },
   \phi_u =
    arg \left( y_u E- { x_u x'_u B^2\over A }\right),\cr
    S_{D_{L1}} &= {z_d C\over E'_d},\;
    E'_d = \abs{ y_d E - {x_dx'_d B^2\over A} },\;
   \phi_d =
arg \left( y_dE - {x_dx'_d B^2\over A }\right),\cr
S_{U_{L2}} &= {x_u B\over A},\cr
S_{D_{L2}} &= {x_d B\over A},\cr}
$$
we can see that the $W$'s and $V_{KM}$ do have similar hierarchical patterns,
but have different Clebsch factors associated with their entries.

When a specific model is given, one can calculate all the Clebsch factors and
make some definite predictions for that particular model.
For example, the ADHRS Model 6, which gives results in good agreement with
the experimental data, has the following four effective fermion mass operators
$$
\eqalignno{
O_{33} &= 16_3\;  10\;  16_3,\cr
O_{23} &= 16_2\; {A_Y\over A_X}\;  10\;  {A_Y\over A_X} 16_3,\cr
O_{23} &= 16_2  {A_X\over M}  \;  10\;  {A_{B-L}\over A_X}\; 16_2
              \;\;\;   {\hbox{or other 5 choices}},\cr
O_{12} &= 16_1 \left( {A_X\over M }\right)^3\; 10\;
 \left( {A_X \over M}\right)^3
          16_2,&(7.17)
\cr}
$$
where $A_X, A_Y, A_{B-L}$ are adjoint's of $SO(10)$ with
VEV's in the $SU(5)$
singlet, hypercharge, and $B-L$ directions.
There are six choices of $O_{22}$ operators
which give the same predictions for
the fermion masses and mixings, but
different Clebsches for other operators
appearing above $M_G$.
Fortunately, they do not enter the  leading
terms of the most important mixing
matrix elements $W_{E_{L32}}, W_{E_{L31}}, W_{E_{R32}}, W_{E_{R31}}$,
 which appear in
the leading contributions to the amplitudes of
LFV processes and the electric
dipole moment.

The magnitude of the mixing matrix elements $V_{KM 32}, V_{KM 31}, W_{E_{L32}},
W_{E_{L31}},\\ W_{E_{R32}},
 W_{E_{R31}}$, and the relevant Clebsch factors are listed in
 Tables 2 and 3.

\begin{center}
{\large\bf{Table 2}}

\vskip .20in
\begin{tabular}{|c|c|c|c|c|c|c|}
\hline
&$u$ & $d$ & $e$ &$eu$ & $q\ell$ &$ n\ell$\cr
\hline
$x$& $-1$ & $-{1\over 6}$ & ${3\over 2}$ & $-6$ & ${1\over 4}$ & 0\cr
\hline
$x'$ & $-1$ & $-{1\over 6}$ & ${3\over 2}$ & $-6$ & ${1\over 4}$ & 0\cr
\hline
$y$ & 0 & 1 & 3 & - & -& -\cr
\hline
$z$ & $-{1\over 27}$ & 1 & 1 & $-{ 1\over 27}$ & 1& 125\cr
\hline
$z'$&$-{1\over 27}$ & 1 & 1 & $-{1\over 27}$ & 1 & 125\cr
\hline
\multicolumn{7}{|c|}{$\widehat{x'_e}
= {9\over 20},\; \overline{x_e} =- 3$}\cr
\hline
\end{tabular}
\end{center}

{\bf Table 2:} Clebsch factors for Yukawa coupling matrices in
ADHRS \mbox{model 6}.
\newpage

\medskip

\centerline{\bf{Table 3}}

\vskip .20in

\begin{center}
\begin{tabular}{|c|c|c|c|}
\hline
&ADHRS models&Model 6&Relevant process\cr
\hline
$\abs{W_{E_{L_{32}}}/V_{ts} }$&
$\abs{ {\wh{x}'_e - x'_e\over x_d - x_u} }$ &
 1.26&\cr
&&&\cr
\hline
$\abs{W_{E_{R32}}/V_{ts} }$ &
$\abs{ {\bar{x}_e -x_e\over x_d - x_u} }$
& 5.4 &\cr
&&&\cr
\hline
$\abs{ W_{E_{L31}}/V_{td} }$ & $\abs{ {z'_e y_d (\wh{x}'_e - x'_e)\over
  z_d y_e (x_d - x_u)} }$ & 0.42 &\cr
&&&\cr
\hline
$\abs{ W_{E_{R31}}/V_{td} }$ & $\abs{ {z_e y_d (\bar{x}_e - x_e)
    \over z_d y_e (x_d - x_u)} }$& 1.8&\cr
&&&\cr
\hline
$\abs{ { W_{E_{L32}}W_{E_{R31}}\over V_{ts} V_{td} } }$ &
$\abs{ {z_e y_d(\bar{x}_e-x_e)(\wh{x}'_e - x'_e)
        \over z_d y_e (x_d-x_u)^2 }}$ &
        2.268 &  $\mu\to e\gamma$ amplitude\cr
&&&\cr
\hline
$\abs{ {W_{E_{R32}} W_{E_{L31}}\over V_{ts}V_{td} } }$ &
$\abs{ {z'_e y_d (\bar{x}_e - x_e)(\wh{x}'_e-x'_e)\over
      z_d y_e (x_d - x_u)^2 } } $& 2.268 &$\mu \to e\gamma$ amplitude\cr
&&&\cr
\hline
$
\Bigg|
\left(
{
\sqrt{2} W_{E_{L31}} W_{E_{R32}}\over
\sqrt{ | W_{E_{L32}} W_{E_{R31}}|^2 +
|W_{E_{R32}} W_{E_{L31}} |^2}
}
\right) /
 \left( { V_{td}\over V_{ts} }\right)
\Bigg|$
& $ \bigg| {
\sqrt{2} z'_e z_e y_d\over \sqrt{ (z^2_e + z'^2_e)} z_d y_e
}\bigg| $ &
${1\over 3} $ & $d_e$\cr
\hline
\end{tabular}
\end{center}

{\bf Table 3:} The relevant Clebsch factors for $\mu \to e\gamma$
and $d_e$ in ADHRS \mbox{model 6}.

\vskip .4in

In ADHRS models $\tan\beta$ is large.
The $\mu \to e\gamma$ rate for large $\tan\beta$ has been calculated
in Sec.\ V and VI for $W_{E_{L32}} = W_{E_{R32}} = V_{ts}$
and $W_{E_{L31}} = W_{E_{R31}} = V_{td}$.
To obtain the predictions of
ADHRS models we only have to multiply the results
by the suitable Clebsch factors.
The relevant Clebsch factors for Model 6 are listed in Table 3.
For a generic realistic GUT model with small tan$\beta$, for example the
modified ADHRS models in which
the down type Higgs lies predominantly in some
fields which do not interact with the three low energy
generations and contain
only a small fraction of the
doublets in the 10 which interact with the low
energy generations \cite{CDW}, most of analysis should still hold.
In this case the leading contributions to $\mu \to e\gamma$ are
the same ones as in the minimal $SO(10)$ model of Ref. \cite{BHS}
(Fig.\ 10
$b_{L,R}, c_{L,R},
c'_{L,R}$ of \cite{BHS}).
The diagrams $c_{LR}, c'_{LR}$ involve
the corrections to the trilinear scalar
couplings.

In the one-loop approximation the
leading corrections to ${\boldzeta}_E$ at
$M_G$ contain pieces proportional to
 ${\boldlambda}_E, {\boldlambda}_E ({\boldlambda}^\dagger_E
{\boldlambda}_E + 3 {\boldlambda}^\dagger_{q\ell} {\boldlambda}_{q\ell} +
 {\boldlambda}^\dagger_{n\ell}
{\boldlambda}_{n\ell}), (2 {\boldlambda}_{E} {\boldlambda}^\dagger_{E}+ 3
 {\boldlambda}_{eu}
{\boldlambda}^\dagger_{eu}){\boldlambda}_E$
respectively.
The piece proportional to ${\boldlambda}_E$ can be absorbed into
${\boldzeta}_{E_0}$ by a
redefinition of $A_E$, the other two
pieces are proportional to the product of
${\boldlambda}_E$ and the corrections to the scalar masses,
$$
\eqalignno{
\Delta{\boldzeta}_E     &= \Delta{\boldzeta}_{E_R}
+ \Delta{\boldzeta}_{E_L}\cr
\Delta{\boldzeta}_{E_R} &= {1\over \mu_{E_R}} \Delta {\bf{m}}_E^2
\;{\boldlambda}_E\cr
\Delta{\boldzeta}_{E_L} &= {1\over \mu_{E_L}} {\boldlambda}_E
\;\Delta{\bf{m}}_{L}^2&(7.18)\cr}
$$
where $\mu_{E_{L}}, \mu_{E_{R}}$ are proportional constants
$(\mu_{E_R} = \mu_{E_L} = {6m^2_0 + A^2_0 \over 3A_0}$
in one-loop approximation).
The LFV couplings in Fig.\ 2e,
 $\: \wt{e}^T_R U^T_E \Delta
{\boldzeta }_E U_L \wt{e}_L v_D$, now can be
written as
$$
\eqalignno{
&{1\over \mu_{E_R}} \wt{e}^T_R U^T_E \Delta {\bf{m}}^2_E
    {\boldlambda}_E U_L \wt{e}_L v_D + {1\over \mu_{E_L}}
    \wt{e}^T_R U^T_E {\boldlambda}_E
    \Delta{\bf{m}}^2_L U_L \wt{e}_L v_D\cr
= & {1\over \mu_{E_R}} \wt{e}_R^T \Delta{\overline{\bf{m}}}^2_E W^*_{E_R}
    \bar{\boldlambda}_E W^\dagger_{E_L}
       \wt{e}_L v_D
    + {1\over \mu_{E_L}} \wt{e}^T_R W^*_{E_R} \bar{\boldlambda}_E
    W^\dagger_{E_L} \Delta{\overline{\bf{m}}}^2_L
    \wt{e}_L v_D,&(7.19)\cr}
$$
where the overline means that the matrix is diagonal.
Again, the amplitudes are given by the same formulas as in \cite{BHS}
(eqn.\ 29,  30), except that $V^e_{32}V^e_{31}(V^{e*}_{33})^2$
has to be replaced by $W_{E_{L32}} W_{E_{R31}} W^*_{E_{L33}} W^*_{E_{R33}}$,
and $W_{E_{R32}} W_{E_{L31}}W^*_{E_{R33}}W^*_{E_{L33}}$, and ${5\over 7} I'_G$
by ${1\over \mu_{E_{R}}} {\Delta}\overline{m}^2_{E33}$ and ${1\over \mu_{E_L}}
{\Delta}\overline{m}^2_{L33}$.
The   results in \cite{BHS} are only modified by some
multiplicative factors and therefore represent the
central values for the LFV processes.

It was pointed out in \cite{BHS,DH}
that the electric dipole moment of
the electron $(d_e)$ constitutes an independent and equally
important signature for the $SO(10)$ unified theory as $\mu \to e \gamma$
 does.
The diagrams which contribute to the electric dipole moment of the
electron are the same as the ones which contribute to $\mu \to e\gamma$,
 with $\mu_L(\mu^c_L)$ replaced by $e_L(e_L^c)$.
Thus a simple relation between $d_e$ and
the $\mu\to e\gamma$ rate was obtained in
the minimal $SO(10)$ model \cite{BHS},
$$
\Gamma (\mu \to e\gamma) = {\alpha\over 2} m^3_\mu |F_2|^2,\eqno(7.20)
$$
$$
|d_e| = e|F_2| \abs{ {V_{td}\over V_{ts}}} \sin\phi = e \sqrt{
{2\Gamma (\mu\to e\gamma)\over \alpha m^3_\mu}}\abs{
{V_{td}\over V_{ts} } } \sin\phi,\eqno(7.21)
$$
where $\phi$ is an unknown new CP violating phase defined by
$$
Im [m_\tau(V^e_{31})^2 (V^{e*}_{33})^2] = |m_\tau(V^e_{31})^2
(V^{e*}_{33})^2| \sin \phi.
$$
In a more generic $SO(10)$ model, such as the ADHRS model,
 we still have
this simple relation but the mixing matrix elements have to be
replaced by the $W$'s:
$$
|d_e| = e \sqrt{ {2\Gamma (\mu \to e\gamma)\over \alpha m^3_\mu} }
{ \sqrt{2}|W_{E_{L31}}W_{E_{R31}}|\over \sqrt{
|W_{E_{L32}} W_{E_{R31}}|^2 + |W_{E_{R31}}W_{E_{L31}}|^2} }
 \sin\phi',\eqno(7.22)
$$
where $\phi'$ is defined by
$$
 Im[m_\tau W_{E_{L31}}W_{E_{R31}} W^*_{E_{R33}}W^*_{E_{L33}}]=
|m_\tau W_{E_{L31}}W_{E_{R31}}W^*_{E_{L33}}W^*_{E_{R33}}| \sin \phi'.
$$
In particular, in ADHRS models there is only one CP violating phase, so the
phase $\phi'$
can be related to the phase appeared in the KM matrix of the Standard Model.
{}From eqn. (7.13) (7.14) (7.16) we can see that
$\phi' \approx \phi_e, \phi_e \approx \phi_d \approx
\wt{\phi}, \phi_u =0$
(because $y_u =0$).
The rephrase invariant quantity $J$ of the KM matrix is given by
$$
\eqalignno{
J &= Im V_{ud} V_{tb} V^*_{td} V^*_{ub}\cr
&\simeq - S_{U_{L1}} S_{D_{L1}} (S_{D_{L2}}
- S_{U_{L2}})^2\sin \phi_d.&(7.23)\cr}
$$
Therefore the CP violating
phase appeared in $d_e$ related to the CP
violation in the Standard Model by
$$
\sin \phi' \simeq {J\over |V_{td}| |V_{ub}|}.\eqno(7.24)
$$

Finally, as  mentioned in the Sec.\ III,
we consider the possibility that
the slight non-degeneracy
between the first two generation scalar masses could
 give a significant contribution to the flavor
changing processes because of the larger mixing matrix elements.
We still use ADHRS models as an example to estimate this contribution to the
LFV process $\mu \to e\gamma$.
For an order of magnitude estimate, the mass insertion approximation in the
super-KM basis employed in \cite{BH}
will serve as a convenient method.
After rotating the $\Delta {\bf{m}}^2_E$ in
eqn.\ (7.10) to the charged lepton
 mass eigenstate basis,
 the contribution
from the first two generations to ${\Delta m^2_{E_{21}}}$ is
$$
\eqalignno{
{\Delta}{m}^2_{E_{21}} (2{\hbox{-}}1)
\simeq V_{E_{R22}} V^*_{E_{R21}} \Delta m^2_{E22}
    + V_{E_{R12}}
    V^*_{E_{R11}} \Delta m^2_{E_{11}}
    + V_{E_{R22}} V^*_{E_{R11}}
\Delta m^2_{E_{21}}\cr
 \simeq [- {z_e C\over E'_e} { \overline{z'^2_e} C^2 +
\overline{{y}^2_e}
    |E|^2 + \overline{x^2_e} B^2\over A^2}  + {z_eC\over E'_e}
    {\overline{z^2_e}  C^2\over A^2} +
    e^{-i\phi_e} {\overline{z_e {y}_e}} {CE\over A^2} ]
    \Delta m^2_{E_{33}}\cr
\simeq - {z_e\over y_e} \left[ {\overline{y^2_e} C|E|
    + \overline{x^2_e} {CB^2\over |E|}
           + \overline{z_ey_e}C|E|\over A^2}
            \right]
    \Delta m^2_{E_{33}} \hspace{.7in}\cr
{(\hbox{assume }} \; z_e = z'_e \;
    {\hbox{as in ADHRS model}}).
&(7.25)
\cr}
$$
Compared with the result found in \cite{BH} for minimal $SU(5)$:
$$
\eqalignno{
\Delta m^2_{E_{21}} (BH)
&=      V^*_{ts} V_{td} \Delta m^2_{E_{33}} \cr
&\simeq - {z_d C\over E'_d}
{(x_d - x_u)^2B^2\over A^2} \Delta m^2_{E_{33}}\cr
&\simeq - {z_d\over y_d} {(x_d - x_u)^2 {CB^2\over |E|}\over A^2}
\Delta m^2_{E_{33}}, &(7.26)
\cr}
$$
we can see that if the Clebsch factors
are $O(1)$, this contribution is comparable
to that of the minimal $SU(5)$ model.
In order for this contribution to be
competitive with the dominant diagrams (Fig.\ 10
$b_{L,R}, c_{L,R}, c'_{L,R} $ of \cite{BHS})
which are enhanced by ${m_\tau\over m_\mu}$, large
Clebsch factors are required. While it is possible to have large Clebsch
factors, we consider
them as model dependent, not generic to all realistic unified theories.

\newpage
\noindent{\bf  VIII. Conclusions}

In supersymmetric theories,
the Yukawa interactions which violate flavor symmetries
not only generate the quark and lepton mass matrices, but necessarily also lead
to radiative breaking of flavor symmetries in the squark and slepton mass
matrices, leading to a variety of flavor signals. While such effects have been
well studied in the MSSM and, more recently, in minimal unified models, the
purpose of this paper
has been to explore these phenomena in a wide class of grand
unified models which have realistic fermion masses.

We have argued that, if the hardness scale $\Lambda_H$
is above $M_G$,
the expectation for all realistic grand unified
supersymmetric models is that non-trivial flavor mixing matrices should occur
at {\it all} neutral gaugino vertices.
These additional, weak scale, flavor violations
are expected to have a form similar to the Kobayashi-Maskawa matrix. However,
the precise values of the matrix elements are model dependent and have
renormalization group scalings which differ
from those of the Kobayashi-Maskawa
matrix elements.

It is the non-triviality of the flavor mixing matrices of neutral gaugino
couplings in the up quark sector which strongly distinguishes between the
general and minimal unified models, as shown in Table 1. Although the minimal
unified models provide a simple approximation to flavor physics, they are not
realistic, so we stress the important new result that flavor mixing in the up
sector couplings of neutral gauginos is a necessity in unified models. this
leads to four important phenomenological consequences. While the $D^0-
\bar{D}^0$
mixing induced by this new flavor mixing is generally not close to the
present experimental limit, it could be much larger than that predicted in the
standard model.

The new mixing in the up-quark sector implies that there may be significant
radiative contributions to the up quark mass matrix which arise when the
superpartners are integrated out of the theory. This is illustrated in Figure
4, where the new mixing matrix elements have been taken to be a factor of three
larger than the corresponding Kobayashi-Maskawa matrix elements. In this case
the entire up quark mass could be generated by such a radiative mechanism:
above the weak scale the violation of up quark flavor symmetries lies in the
squark mass matrix.

The electric dipole moment of the neutron, $d_n$, is a powerful probe of the
neutral gaugino flavor mixing induced by unified theories. In the minimal
$SO(10)$ theory, $d_n$ arises from the flavor mixing in the down sector, which
leads to a down quark dipole moment, $d_d$. However, in realistic models the
flavor mixing in the up quark sector leads to a $d_u$ which typically provides
the dominant contribution to $d_n$. Thus the neutron electric dipole moment is
a more powerful probe of unified supersymmetric theories than previously
realized.

The presence of flavor mixing in the up sector plays a very important role in
determining the branching ratio for a proton to decay to $K^0 \mu^+$. In the
minimal models, without such mixings, this branching ratio is expected to be
about $10^{-3}$: the charged lepton mode will not be seen and experimental
efforts must concentrate on the mode containing a neutrino, $K^+ \nu$.
However, including these mixings the charged lepton branching ratio is greatly
increased to about $0.1$. While this number is very model dependent, we
nevertheless think that this effect greatly changes the importance of searching
for the charged lepton mode.

These four phenomenologocal consequences are sufficiently interesting that we
stress once more that they appear as a necessity in a wide class of unified
theories. The absence of mixing in the up sector is a special feature
of the minimal models. Since the flavor sectors of the minimal models
must be augmented to obtain realistic fermion masses, any conclusions based on
the absence of flavor mixings in the up sector are specious.

A second topic addressed in this paper
is the effect of large $\tan \beta$ on the lepton process, $\mu \rightarrow e
\gamma$ which is expected in unified supersymmetric $SO(10)$ models. The
amplitude for this process has a contribution proportional to $\tan \beta$.
In this paper, we have found that the naive
expectation that large $\tan \beta$ in supersymmetric $SO(10)$ is excluded
by  $\mu \rightarrow e \gamma$ is incorrect, at least for all values of the
superpartner masses of interest. Contour plots
for the  $\mu \rightarrow e \gamma$
branching ratio are shown in Figures 7 and 8. It depends sensitively on the
parameter $\Delta$, which is the mass splitting between the scalar electron
and scalar tau, and is plotted in Figure 9. Lower
values of the top quark Yukawa coupling, which for large $\tan \beta$ still
give allowed predictions for the $b/\tau$ mass ratio, give a much reduced value
for $\Delta$, thereby reducing the  $\mu \rightarrow e \gamma$ rate and
partially compensating the $\tan^2 \beta$ enhancement. A further significant
suppression of an order of magnitude is induced by the renormalization group
scaling of the leptonic flavor mixing angles, and is shown in Figure 10. The
net effect is that while the case of $\tan \beta \approx m_t/m_b$ is not
excluded in SO(10), the  $\mu \rightarrow e \gamma$ rate is still typically
larger than for moderate $\tan \beta$, so that this process
provides a more powerful probe of the theory as $\tan \beta $ increases.

For large $\tan \beta$,  $\mu$ and $M_2$ become the physical masses of the two
charginos. The  $\mu \rightarrow e \gamma$ contours of Figure 8
show that $\mu$ and $M_2$ should not be too large, providing an important limit
to the chargino masses in the large $\tan \beta$ limit. Furthermore, this
constrains the LSP mass to be quite small.
We find that in this region it is still possible for the LSP to account
for the observed dark matter, and even to critically close the universe, as can
be seen from Figure 11.
However, the requirement that the LSP mass be larger than 45 GeV suggests that
the two light charginos will not be light enough to be discovered at LEP II.

As an example of theories with both a realistic flavor sector and large $\tan
\beta$ we studied the models introduced by Anderson et al. The flavor sectors
of these theories are economical: the free parameters can all be fixed from the
known quark and lepton masses and mixings. Hence the flavor mixing matrices at
all neutral gaugino vertices can be calculated. These are shown for the lepton
sector of model 6 in Table 3. The Clebsch factors enhance the $\mu \rightarrow
e \gamma$ amplitude by a factor of 2.3, and suppress $d_e$ by a factor of 3.
Even taking the top quark Yukawa coupling to have its lowest value the rate for
$\mu \rightarrow e \gamma$ in this theory is very large. Another interesting
feature of these theories is that the flavor sectors contain just a single CP
violating phase. This means that the phase which appears in the result for
$d_n$ and $d_e$ can be computed: since it is closely related to the phase of
the Kobayashi-Maskawa matrix it is not very small. That which appears in
$d_e$ is given in eqn.\ (7.24) and is numerically
about 0.2. We have computed the
radiative corrections to $m_u$ in the ADHRS models and have found that the new
mixing matrices in the up sector are not large enough to yield sizable
contributions: thus the ADHRS analysis of the quark mass matrices is not
modified. Furthermore, due to a cancellation special to these theories, there
is no contribution to $d_n$ from the up quark at one loop.

\newpage
\noindent {\bf Acknowledgements}

    The authors would like to thank Hitoshi Murayama for many useful
discussions. This work was supported in part by the Director, Office of Energy
Research, Office of High Energy and Nuclear Physics, Division of High Energy
Physics of the U.S. Department of Energy under contract No. DE-AC03-76SF00098
and in part by the National Science Foundation under Grant No. PHY-90-21139.
The work of N.A-H was supported by an NSERC '67 fellowship.

\vskip .6in

\noindent    Note Added:

While finalizing this work,
we received a preprint by Ciafaloni,
Romanio and Strumia\cite{CRS},
where the large $\tan\beta$ scenario is also considered. However,
unlike this work, they assume strict universality in soft scalar masses,
such that imposing electroweak symmetry breaking leads
them into a region of
parameter space with
a high mass (1 TeV) for the sleptons.
In their discussion of general models, they do not include
flavor violating RG scaling of scalar masses above $M_G$.

\newpage

\noindent{\bf{Appendix}}

In this appendix, we first give a more complete treatment of mixing
matrix scaling in the lepton sector, and then give a treatment
for the quark sector.

Let us  return to (5.7) and consider the effect of including the
$({\boldzeta}^\dagger_E{\boldzeta}_E)_{3i}$
term. In general the scaling from $M_{PL}$ to $M_G$
will generate a
${\boldzeta}^\dagger_E{\boldzeta}_E$
not diagonal in the same basis as ${\boldlambda}^\dagger_E
{\boldlambda}_E$,
  so we
expect some non-zero $({\boldzeta}^\dagger_{E}{\boldzeta}_E)_{3i}$.
{}From the RGE for ${\boldzeta}_E$, neglecting gauge couplings,
$$
- {d\over dt} {\boldzeta}_E = {\boldzeta}_E [ 5{\boldlambda}^\dagger_E
{\boldlambda}_E
+ Tr (3{\boldlambda}^\dagger_D
{\boldlambda}_D + {\boldlambda}^\dagger_E {\boldlambda}_E)]
+ {\boldlambda}_E [4{\boldlambda}^\dagger_E {\boldzeta}_E + Tr
(6{\boldzeta}_D {\boldlambda}_D^\dagger + 2
{\boldzeta}_E{\boldlambda}_E^ \dagger)].\eqno(A.1)
$$
We have
$$
\eqalignno{
-{d\over dt} ({\boldzeta}^\dagger_E{\boldzeta}_E) &=5[{\boldzeta}^\dagger_E
{\boldzeta}_E
{\boldlambda}^\dagger_E{\boldlambda}_E
+ {\boldlambda}^\dagger_E{\boldlambda}_E{\boldzeta}^\dagger_E
{\boldzeta}_E]+ 2Tr(3{\boldlambda}^\dagger_D
{\boldlambda}_D + {\boldlambda}^\dagger_E {\boldlambda}_E)
{\boldzeta}^\dagger_E{\boldzeta}_E\cr
&+ 8{\boldzeta}^\dagger_E {\boldlambda}_E{\boldlambda}^\dagger_E{\boldzeta}_E
+ ({\boldzeta}^\dagger_E {\boldlambda}_E
+ {\boldlambda}_E^\dagger{\boldzeta}_E)
Tr(6{\boldzeta}_D{\boldlambda}^\dagger_D
+ 2{\boldzeta}_E{\boldlambda}^\dagger_E).&(A.2)
\cr}
$$
Then, to first order in the off diagonal parts of
 ${\boldzeta}^\dagger_E{\boldzeta}_E$ and
${\boldzeta}_E{\boldzeta}_E^\dagger$, and keeping only third
 generation Yukawa couplings we have
$$
- {d\over dt} ({\boldzeta}^\dagger_E {\boldzeta}_E)_{3i}
= ({\boldzeta}^\dagger_E{\boldzeta}_E)_{3i}[17 {\lambda}^2_\tau + 6
{\lambda}^2_b + 6\eta {\lambda}_b {\lambda}_\tau],\eqno(A.3)
$$
where $\eta \equiv {\zeta_{D_{33}}\over \zeta_{E_{33}} }$.
Because of the large numerical coefficient in front of
$\lambda^2_\tau, \lambda^2_b$
in the above equation,
(${\boldzeta}^\dagger_E{\boldzeta}_E)_{3i}$ is driven to zero more
rapidly than $W_{L3i}$, after which it ceases to have any effect on the
running of $W_{L3i}$.
More explicitly, from (5.7) we have that
$$
{d\over dt} (m^2_{L3i}(t)e^{\int^t_0 dt'\lambda^2_\tau (t')}) =- 2
    ({\boldzeta}^\dagger_E{\boldzeta}_E)_{3i}(t)
e^{\int^t_0 dt' \lambda^2_\tau(t')}.
\eqno(A.4)
$$
Solving (A.3) for $({\boldzeta}^\dagger_E{\boldzeta}_E)_{3i}(t)$ and
 inserting into (A.4) we get
$$
-{d\over dt} \left( m^2_{L 3i}(t)e^{\int^t_0 dt'\lambda^2_\tau (t')}\right)
= 2({\boldzeta}^\dagger_E{\boldzeta}_E)_{3i}(M_G)e^{-\int^t_0
 dt'[16 \lambda^2_\tau
+6\lambda^2_b + 6\eta \lambda_b \lambda_t](t')}.\eqno(A.5)
$$
Integrating (A.5), we find
$$
\eqalignno{
m^2_{L 3i} (M_S)e^{I_\tau} - m_{L 3i}^2 (M_G) &=
-2 \int_0^{ {1\over 16\pi^2}\log
{M_G\over M_S}} dt
e^{-\int^t_0 dt'[16\lambda^2_\tau + 6 \lambda^2_b + 6\eta
 \lambda_b\lambda_\tau](t')}\cr
&\times ({\boldzeta}^\dagger_E{\boldzeta}_E)_{3i}(M_G)\cr
&\equiv \delta ({\boldzeta}^\dagger_E{\boldzeta}_E)_{3i} (M_G).&(A.6)\cr}
$$
So, we have
$$
m^2_{L 3i}(M_S) = e^{-I_\tau}[m^2_{L 3i} (M_G) + \delta
({\boldzeta}^\dagger_E {\boldzeta}_E)
(M_G)].\eqno(A.7)
$$
We expect $m^2_{L 3i}$ and
${({\boldzeta}^\dagger_E {\boldzeta}_{E})}_{3i}$
 to be related by some
combination of Clebsches $x$ at $M_G$ as follows:
$$
({\boldzeta}^\dagger_E{\boldzeta}_E)_{3i} =
 {A^2_0\over m^2_0} xm^2_{L 3i}\eqno(A.8)
$$
Where $A_0, m^2_0$ are the universal $A$ parameter and scalar mass
at $M_{PL}$,
respectively.
Then, we have from (A.7)
$$
W^\dagger_{L33}
W_{L 3i} (M_S) = e^{-I_\tau}{\Delta m^2(M_G)\over \Delta m^2 (M_S)}
[1 + \delta {A^2_0\over m^2_0} x] W^\dagger_{L33}W_{L 3i} (M_G).\eqno(A.9)
$$
Clearly if $\delta {A^2_0\over m^2_0} x$ $\ll 1$,
inclusion of the $({\boldzeta}^\dagger_E{\boldzeta}_E)_{3i}$ term in (5.7)
do not change any of our results.
If $\delta{A^2_0\over m^2_0} x$  $\sim 1 $ or $\gg 1$, we
can still of course use (A.9), but the suppression effect may disappear.
A simple estimate shows, however, that $\delta$ itself is already
small $\sim {1\over 10}$,
and so we are only in trouble if ${A^2_0\over m^2_0} x$ is big.
To see this, replace
$\lambda_\tau, \lambda_b$ and $\eta$ by some average values
$\bar{\lambda}_\tau, \bar{\lambda}_b$ and
$\bar{\eta}$ in the expression (A.6) for $\delta$.
Then,
$$
\eqalignno{
\delta &= -2 \int_0^{ {1\over 16 \pi^2}\log {M_G\over M_S}}
    e^{-t(16\bar{\lambda}_\tau^2
    + 6 \bar{\lambda}^2_b + 6\bar{\lambda}_b
    \bar{\lambda}_t \bar{\eta} )}\cr
&= - {1\over 8\bar{\lambda}^2_t + 3(\bar{\lambda}^2_b
    +\bar{\eta}\bar{\lambda}_b\bar{\lambda}_\tau)}
    \left[ e^{- {1\over 16\pi^2} \log {M_G\over M_S}
    (16 \bar{\lambda}^2_\tau + 6 \bar{\lambda}^2_b +
    6\bar{\eta}\bar{\lambda}_b\bar{\lambda}_\tau)} \right].
&(A.10)
\cr}
$$
So,
$$
|\delta | < {
1\over 8 \bar{\lambda}^2_\tau + 3(\bar{\lambda}^2_b  +\bar{\eta}
\bar{\lambda}_b
\bar{\lambda}_\tau).
}\eqno(A.11)
$$
For the $\bar{\lambda}$'s between 0.5 and 1, and $\bar{\eta} \sim 1, |\delta |$
ranges from ${1\over 3}$ to ${1\over 15}$.

How can we qualitatively
understand the above results for the scaling of mixing
matrices?
The renormalization group equations
try to align the soft supersymmetry breaking flavor
matrices with whatever combination of flavor matrices responsible for
their renormalization.
However, because a given coupling
can only be renormalized by harder couplings, there
is a hierarchy in
which flavor matrices affect the running of others.
The Yukawa matrices,
being dimensionless, can only be affected by other Yukawa
matrices.
In the lepton sector, this is the reason that the basis in which e.g.
${\boldlambda}^\dagger_E{\boldlambda}_E$ is diagonal does not change.
Next, the soft trilinear terms, having mass dimension one,
can only be affected
by other trilinear terms and Yukawa couplings.
Again in the lepton sector this means that e.g.
 ${\boldzeta}^\dagger_E {\boldzeta}_E$
tries to align
itself with ${\boldlambda}^\dagger_E{\boldlambda}_E$.
Finally, the scalar mass, having dimension two,
are affected by everything:
${\bf{m}}^2_L$ tries to align with
${\boldlambda}^\dagger_E{\boldlambda}_E$,
 but suffers
interference from ${\boldzeta}^\dagger_E {\boldzeta}_E$, unless
 ${\boldzeta}^\dagger_E
{\boldzeta}_E$
 is diagonal in the same basis as
${\boldlambda}^\dagger_E{\boldlambda}_E$.
Even if ${\boldzeta}^\dagger_E{\boldzeta}_E$ is not diagonal in the same
basis as
${\boldlambda}^\dagger_E {\boldlambda}_E$, it is trying to align itself with
 ${\boldlambda}^\dagger_E{\boldlambda}_E$,
so ${\bf{m}}^2_L$ will still tend to align with
${\boldlambda}^\dagger_E{\boldlambda}_E$.

{}From the above discussion, it is clear that the situation
is slightly
complicated in the quark sector.
In the lepton sector, there was a fixed direction in flavor space
given by
${\boldlambda}_E$, with which the soft matrices aligned.
In the quark sector, we have both ${\boldlambda}_U$ and ${\boldlambda}_D$, and
${\boldlambda}_U{\boldlambda}_U^\dagger$, ${\boldlambda}_D
{\boldlambda}_D^\dagger$
 are misaligned $(V_{KM}\neq 1)$.
This complicates the analysis for $W_{U_L}, W_{D_L}$ so we discuss them last.
Let us now examine the scaling of $W_{U_R}, W_{D_R}$.
(Throughout the following, we assume degeneracy between first two generation
scalar masses,
we neglect all Yukawa coupling matrix
eigenvalues except those of the
third generation, and we do not
include the effect of trilinear soft terms in the scaling.
The last assumption is made for simplicity; we can make similar arguments
about the importance of these neglected trilinear terms as
we did above in the lepton sector.)

First, we show that the basis in which
${\boldlambda}^\dagger_U{\boldlambda}_U$ is diagonal remains
fixed.
The RGE for ${\boldlambda}^\dagger_U{\boldlambda}_U$ is
$$
- {d\over dt}{\boldlambda}^\dagger_U{\boldlambda}_U =
6({\boldlambda}^\dagger_U
{\boldlambda}_U)^2
+ 2{\boldlambda}^\dagger_U{\boldlambda}_D
{\boldlambda}^\dagger_D{\boldlambda}_U +
2(3Tr{\boldlambda}_U{\boldlambda}^\dagger_U
- {16\over 3} g^2_3 -3 g^2_2 - {13\over 15}
g^2_1) {\boldlambda}_U^\dagger {\boldlambda}_U.\eqno(A.12)
$$
Working in a basis where ${\boldlambda}^\dagger_U{\boldlambda}_U$ is diagonal,
let us see if
${d\over dt} {\boldlambda}^\dagger_U{\boldlambda}_U$
has off-diagonal components.
We have, (recalling that in this basis
 ${\boldlambda}^\dagger_D{\boldlambda}_D = V_{KM}\bar{{\boldlambda}}^2_D
V^\dagger_{KM})$,
$$
\eqalignno{
- {d\over dt} ({\boldlambda}^\dagger_U
{\boldlambda}_U)_{\stackrel{ij}{i\neq j}}
&=    2 (\bar{{\boldlambda}}_UV_{KM}
\bar{{\boldlambda}}^2_DV^\dagger_{KM}
\bar{{\boldlambda}}_U)_{ij}\cr
&= 2\bar{{\boldlambda}}_{U_i} V_{KM_{i\ell}}
\bar{{\boldlambda}}_{D_{\ell}}^2
V^\dagger_{KM_{\ell j}}
    \bar{{\boldlambda}}_{U_j}\cr
&= 0 \ \hbox{for} \ i, j\neq 3 &(A.13)\cr}
$$
since we neglect all Yukawa's except the third generation.
Similarly, the basis in which ${\boldlambda}^\dagger_D{\boldlambda}_D$
is diagonal does not
change.
Thus, the discussion for the scaling of $W_{U_R}, W_{D_R}$
is completely analogous to that in the lepton sector, and we find
$$
W_{U_{R 3i}}W^\dagger_{U_{R 33}}(M_S) = e^{-2I_t} {\Delta m^2_U(M_G)\over
\Delta m^2_U(M_S)} W_{U_{R 3i}}W^\dagger_{U_{R 33}}(M_G),\eqno(A.14)
$$
$$
W_{D_{R 3i}}W^\dagger_{D_{R 33}}(M_S) = e^{-2I_b} {\Delta m^2_D(M_G)\over
\Delta m^2_D(M_S)} W_{D_{R 3i}}W^\dagger_{D_{R 3}}(M_G).\eqno(A.15)
$$
We now turn to $W_{U_L}, W_{E_L}$.
Let $V^*_{U_L}(t)$ be the matrix diagonalizing ${\boldlambda}_U
{\boldlambda}^\dagger_U(t)$:
$$
{\boldlambda}_U {\boldlambda}^\dagger_U (t)= V^{*}_{U_{L}}(t)
{\bar{{\boldlambda}}}^2_U (t) V^{*\dagger}_{U_{L}} (t).\eqno(A.16)
$$
In the superfield basis in which
${\boldlambda}_U{\boldlambda}^\dagger_U$ is diagonal,
the squark mass matrix is
$\wt{\bf{m}}^{2*}_{Q 3i} = V^\dagger_{U_L}{\bf{m}}^{2*}_Q V_{U_L}$.
Note as before that $\wt{\bf{m}}^{2*}_{Q 3i}
= (W^\dagger_{U_L}\overline{\bf{m}}^{2*}_Q
W_{U_L})_{3i}=
W_{U_{L 3i}}W^\dagger_{U_{L 33}}\Delta m^2_Q$,
so we are interested in ${d\over dt} \wt{\bf{m}}^{2*}_{Q 3i}$. Now,
$$
\eqalignno{
{d\over dt} \wt{\bf{m}}_Q^{2*} = {d\over dt}(V^\dagger_{U_L}
{\bf{m}}^{2*}_QV_{U_L})&=
\left( {d\over dt} V^\dagger_{U_L}\right)
{\bf{m}}^{2*}_Q V_{U_L} + V^\dagger_{U_L} {d\over dt} {\bf{m}}^{2*}_Q V_{U_L}
+ V^\dagger_{U_L}
{\bf{m}}^{2*}_Q {d\over dt}  V_{U_L}\cr
&= \left[ \wt{\bf{m}}^{2*}_Q, V^\dagger_{U_L} {d\over dt} V_{U_L}\right]
    +V^\dagger_{U_L} {d\over dt} {\bf{m}}^{2*}_Q V_{U_L}.&(A.17)\cr}
$$
The second term is the analogue of what we have
already seen in the lepton and right-handed quark sector; using the RGE
for ${\bf{m}}^{2*}_Q$ we find to leading order
$$
\left( V^\dagger_{U_L} {d\over dt} {\bf{m}}^{2*}_Q V_{U_L}\right)_{3i} =-
(\lambda^2_t
    + \lambda^2_b) \wt{\bf{m}}^2_{Q 3i}.\eqno(A.18)
$$
Now, $V^\dagger_{U_L} {d\over dt} V_{U_L}$ is obtained
 from the RGE for ${\boldlambda}_U
{\boldlambda}^\dagger_U$.
Actually, note that
$$
V^\dagger_{U_L} \left( {d\over dt}
{\boldlambda}_U{\boldlambda}^\dagger_U\right)
 V_{U_L} =
\left[ V^\dagger_{U_L} {d\over dt} V_{U_L}, \bar{{\boldlambda}}^2_U\right]
+ {d\over dt}
\bar{{\boldlambda}}^2_U,\eqno(A.19)
$$
so that only $[V^\dagger_{U_L} {d\over dt} V_{U_L}, \bar{{\boldlambda}}^2_U]$
is determined.
(This is a reflection of the fact that $V_{U_L}$ is not unique: let $X (t)$
be any unitary transformation leaving $\overline{\bf{m}}^2_Q(t)$ invariant:
$
\overline{\bf{m}}^2_Q(t) = X^\dagger (t) \overline{\bf{m}}^2_Q(t) X(t).
$
In our case, $X(t)$ is most generally a $U(2)$ matrix in the first
two generation subspace.
Then, if $V_{U_L}$ diagonalizes ${\bf{m}}^{2*}_Q$, so does $V_{U_L}X$.
Under this change, $V^\dagger_{U_L} {d\over dt} V_{U_L}$ is not invariant,
but $[ V^\dagger_{U_L} {d\over dt} V_{U_L},
\bar{{\boldlambda}}^2_U]$ {\it {is}} invariant).
Further, since we neglect first two generations Yukawa eigenvalues,
$[V^\dagger_{U_L} {d\over dt} V_{U_L}, \bar{{\boldlambda}}^2_U]_{ij}
=0 $ for $ i, j= 1,2$,
and only
 $[V^\dagger_{U_L} {d\over dt} V_{U_L}, \bar{{\boldlambda}}^2_U]_{3i (i3)}
= (\mp) \lambda^2_t V^\dagger_{U_L} {d\over dt} V^\dagger_{U_{L i 3
(3 i)}}$
is determined, and we can choose all other components of
$V^\dagger_{U_L} {d\over dt}
V_{U_L}$ to vanish.
{}From the RGE for ${\boldlambda}_U{\boldlambda}_U^\dagger,
$
$$
 \eqalignno{
- {d\over dt} ({\boldlambda}_U{\boldlambda}_U^\dagger) &=
    6({\boldlambda}_U{\boldlambda}^\dagger_U)^2
 + 2(3 Tr {\boldlambda}_U{\boldlambda}_U^\dagger - {16\over 3} g^2_3 - 3g^2_2 -
    { 13\over 15}
    g^2_1) {\boldlambda}_U{\boldlambda}_U^\dagger\cr
&+ \{ {\boldlambda}_U {\boldlambda}^\dagger_U,
{\boldlambda}_D {\boldlambda}^\dagger_D\},&(A.20)\cr}
$$
we find
$$
\eqalignno{
-\left(V^\dagger_{U_L} ({d\over dt}{\boldlambda}_U{\boldlambda}^\dagger_{U})
    V_{U_L}\right)_{3i}
&= \{ \bar{{\boldlambda}}_U^2, V_{KM} \bar{{\boldlambda}}^2_D
V^\dagger_{KM}\}_{3i}\cr
&= {\lambda}^2_t{\lambda}^2_bV_{KM 33}V^\dagger_{KM 3i},
&(A.21)\cr}
$$
and thus
$$
\left( V^\dagger_{U_L} {d\over dt} V_{U_{L 3i}}\right) =- \lambda^2_b
V^\dagger_{KM 3i} V_{KM 33}.\eqno(A.22)
$$
Thus to leading order
$$
[V^\dagger_{U_L} {\bf{m}}^2_Q V_{U_L}, V^\dagger_{U_L}
{d\over dt} V_{U_L}]_{3i}
=- \Delta {{m}}^2_Q \,\lambda^2_b \,V^\dagger_{KM 3i} V_{KM 33},\eqno(A.23)
$$
and finally we have
$$
- {d\over dt} (W_{U_{L3i}} W^\dagger_{U_{L 33}} \Delta m_Q^2) =
(\lambda^2_t + \lambda_b^2) W_{U_{L 3i}}W^\dagger_{U_{L 33}}
\Delta {{m}}^2_Q + \lambda^2_b V^\dagger_{KM 3i} V_{KM 33}\Delta
{{m}}^2_Q.\eqno(A.24)
$$
Similarly we find
$$
- {d\over dt} (W_{D_{L 3i}} W^\dagger_{D_{L 33}}
\Delta {{m}}^2_Q) = (\lambda^2_t +
\lambda^2_b) W_{D_{L 3i}} W^\dagger_{D_{L 33}}
\Delta {{m}}^2_Q
+ \lambda^2_t V^\dagger_{KM 3i}V_{KM 33}\Delta {{m}}^2_{Q}.\eqno(A.25)
$$
We can formally solve the above equations, e.g.
$$
W_{U_{L 3i}} W^\dagger_{U_{L 33}} (M_S) =
e^{-\left( I_t + I_b + \int^{M_G}_{M_S} dt'
\lambda^2_b {V^\dagger_{KM 3i}
V_{KM_{ 33}}\over W_{U_{L 3i}}W^\dagger_{U_{L 33}}}
\right)}
{\Delta {{m}}^2_Q (M_G)\over \Delta {{m}}^2_Q(M_S)}
W_{U_{L 3i}} W^\dagger_{U_{L 33}}
 (M_G),\eqno(A.26)
$$
and, to a good approximation, given that $W_{U_{L 3i}}$ does not
scale very significantly, we can replace
$$
\int^{M_G}_{M_S} dt' \lambda^2_b {V^{\dagger}_{KM 3i}
V_{KM 33}\over W_{U_{L 3i}}W^\dagger_{U_{L 33}} }
\approx I_b {V^\dagger_{KM 3i} V_{KM 33}\over W_{U_{L 3i}}
 W^\dagger_{U_{L 33}} } (M_G).\eqno(A.27)
$$
So, an approximate solution of the RGE for $W_{U_L}, W_{D_L}$ is
$$
W_{U_{L 3i}}
 W^\dagger_{U_{L 33}} (M_S) \approx
e^{ -\left( I_t + I_b \left( 1+
{ V^\dagger_{KM 3i} V_{KM_{ 33}}\over W_{U_{L 3i}}
 W^\dagger_{U_{L 33}}} (M_G)\right) \right)
}
{\Delta {{m}}^2_Q (M_G)\over \Delta {{m}}^2_Q(M_G)}
W_{D_{L 3i}} W^\dagger_{U_{L 33}}
(M_G),\eqno(A.28)
$$
and similarly
$$
W_{D_{L 3i}} W^\dagger_{D_{L 33}}(M_S) \approx
 e^{ -\left( I_b + I_t \left( 1+
{ V^\dagger_{KM 3i} V_{KM 33}
\over W_{U_{L 3i}} W^\dagger_{U_{L 33}}}
(M_G) \right) \right)
}
{\Delta {{m}}^2_Q(M_G)\over \Delta {{m}}^2_Q(M_S)}
 W_{D_{L 3i}} W^\dagger_{D_{L 33}}
(M_G).\eqno(A.29)
$$

The above results are in agreement with qualitative expectations;
the extra terms in the exponential of (A.28) and (A.29) are a
reflection of the fact that the bases in which
${\boldlambda}_U{\boldlambda}^\dagger_U$ and
${\boldlambda}_D{\boldlambda}^\dagger_D$ are diagonal change
with scale. For moderate $\tan\beta$, however, we expect
that the basis in which ${\boldlambda}_U{\boldlambda}^\dagger_U$
is diagonal should not change with scale, and in this
limit the extra term drops out of (A.28).

\newpage

\centerline{\bf Figure Captions}

\medskip

\begin{enumerate}
\item[Fig. 1.] Feynman diagrams contributing to $\mu \to e\gamma$.
\item[Fig. 2] Lepton flavor violating couplings in general
supersymmetric Standard Models.
\item[Fig. 3] Corrections to the
up-type quark mass matrix, proportional to $m_t$.
\item[Fig. 4] Contours for ${\Delta m_u\over m_u}$
in ${m_{\tilde{u}}\over M_{\tilde{g}}} -
{m_{\tilde{t}}\over m_{\tilde{u}}}$ plane,
assuming $m_{\wt{u}_L} = m_{\wt{u}_R} \equiv m_{\wt{u}},\;
m_{\wt{t}_L} = m_{\wt{t}_R}\equiv m_{\wt{t}},\; W_{U_{L31}} = W_{U_{R 31}}
 = {1\over 30},\; {A + \mu \cot \beta\over m_{\tilde{t}} }=3$.
\item[Fig. 5] The diagram which gives the dominant contribution to
 $\mu\to e\gamma$
in the large $\tan\beta$ limit.
A photon is understood to be attached to the diagram in all possible ways.
\item[Fig. 6] The dominant diagram (for $\mu\to e\gamma$) in the
mass insertion approximation.
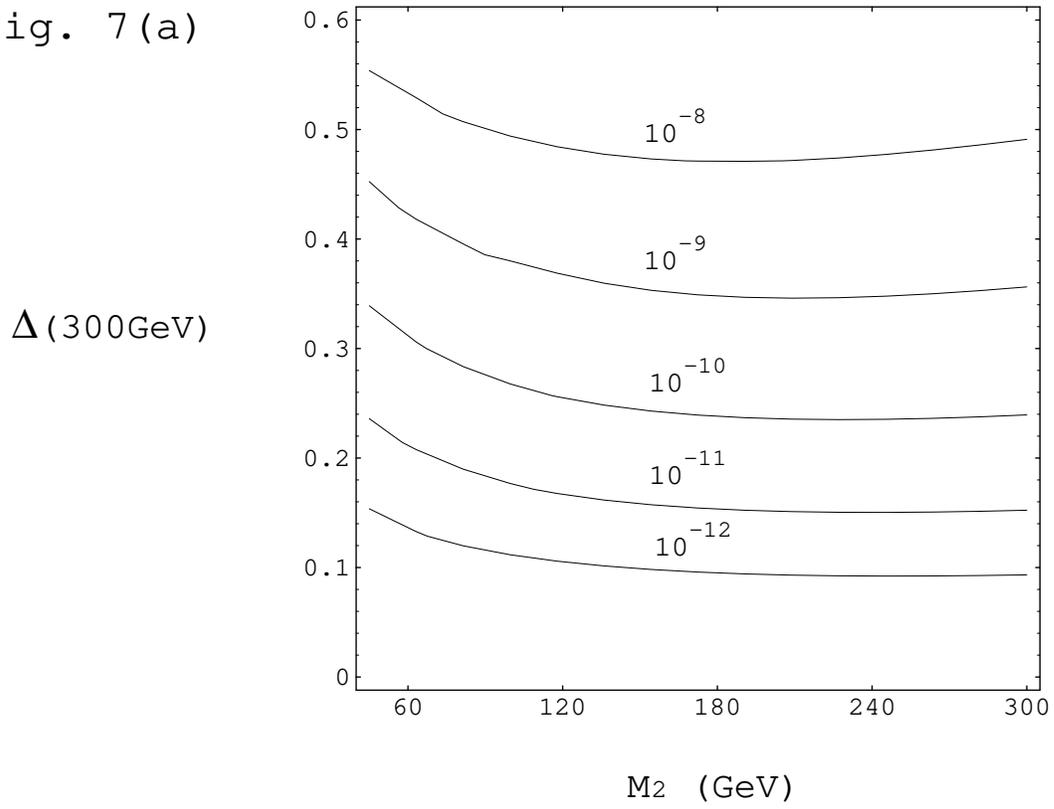
\item[Fig. 7] Contours for $Br(\mu\to e\gamma)$ in $M_2-\Delta$
plane with
$m_{\wt{e}_{L(R)}}= 300 GeV,\; \Delta \equiv m_{\wt{e}_{L(R)}}
 - m_{\wt{t}_{L(R)}},\; W_{E_{L(R) 32}} = 0.04,\;
W_{E_{L(R) 31}}=0.01,$ for (a) $\mu = 100 $ GeV, (b) $\mu = 300$\ GeV.
Contours for negative $\mu$ are virtually identical.
To get $Br(\mu \to e\gamma)$
prediction from a GUT, multiply by appropriate
Clebsch, and $\epsilon$ factor (Fig.\ 10).
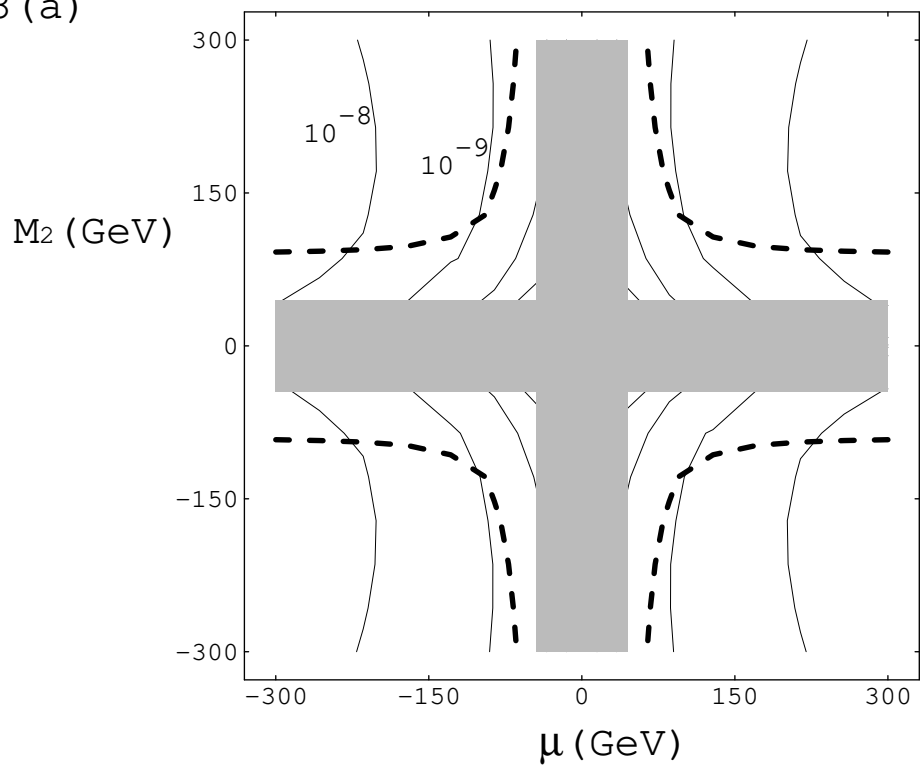
\item[Fig. 8] Contours for $Br(\mu\to e\gamma)$ in $\mu - M_2$ plane for (a) $
\Delta = 0.25,$ (b) $\Delta = 0.5$, with other parameters same as in Fig.\ 7.
The blacked out regions are ruled out by the LEP bound of 45 GeV on
chargino masses. The thick dashed lines are contours for a 45 GeV LSP mass.
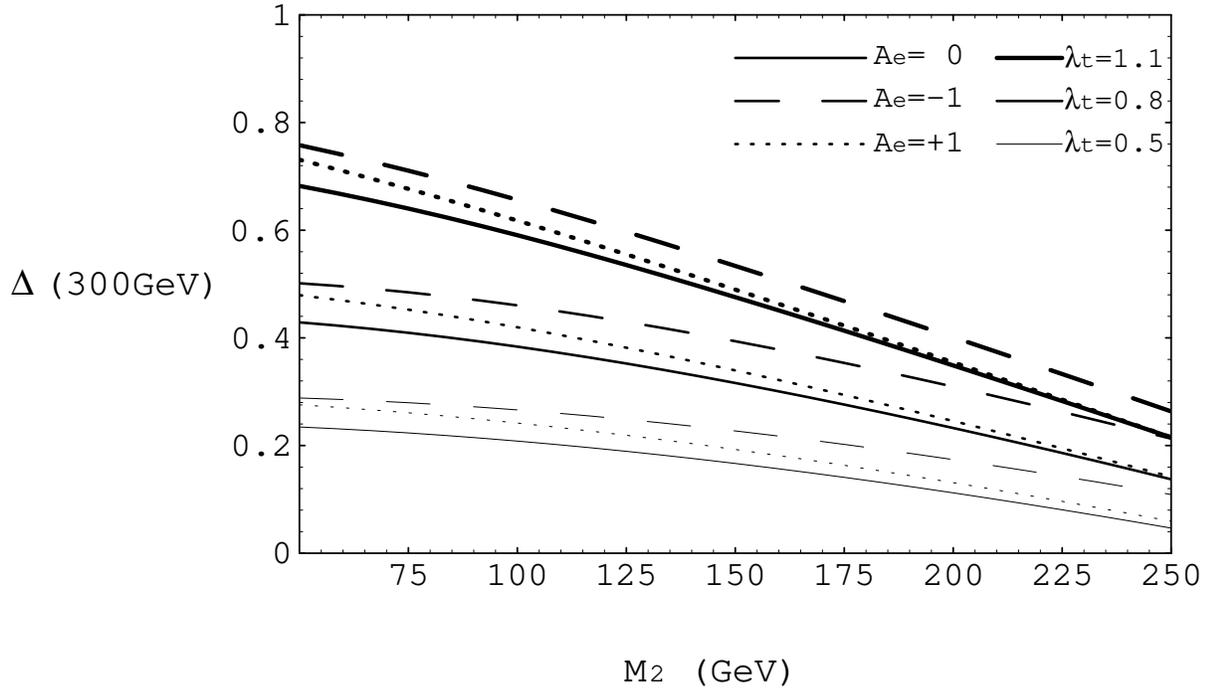
\item[Fig. 9] Plots of the averaged difference between the third and the
first two generations charged slepton masses
$\Delta \equiv {\Delta_L+\Delta_R\over 2},\;
 \Delta_{L(R)}\equiv m_{\wt{e}_{L(R)}}$
(at $M_S$), against $M_2$, for
$\half ({{m}}^2_{\wt{e}_L} +m^2_{\wt{e}_R})=(300\; \mbox{GeV})^2,\;
\lambda_t = \lambda_b=\lambda_\tau$ (at $M_G$) = 0.5, 0.8, 1.1,
$A_e(M_S) = 1, 0, -1$,
two values of the gauge beta function coefficient $b_5$ between $M_G$ and
$M_{PL}$, (a) $b_5 = 3\,$(\mbox{asymptotically free}), (b) $b_5 =- 20$.
Scalar masses are assumed degenerate at $M_{PL} = 2.4 \times 10^{18}$ GeV.
$M_G$ is taken to be 2.7 $\times 10^{16}$ GeV.
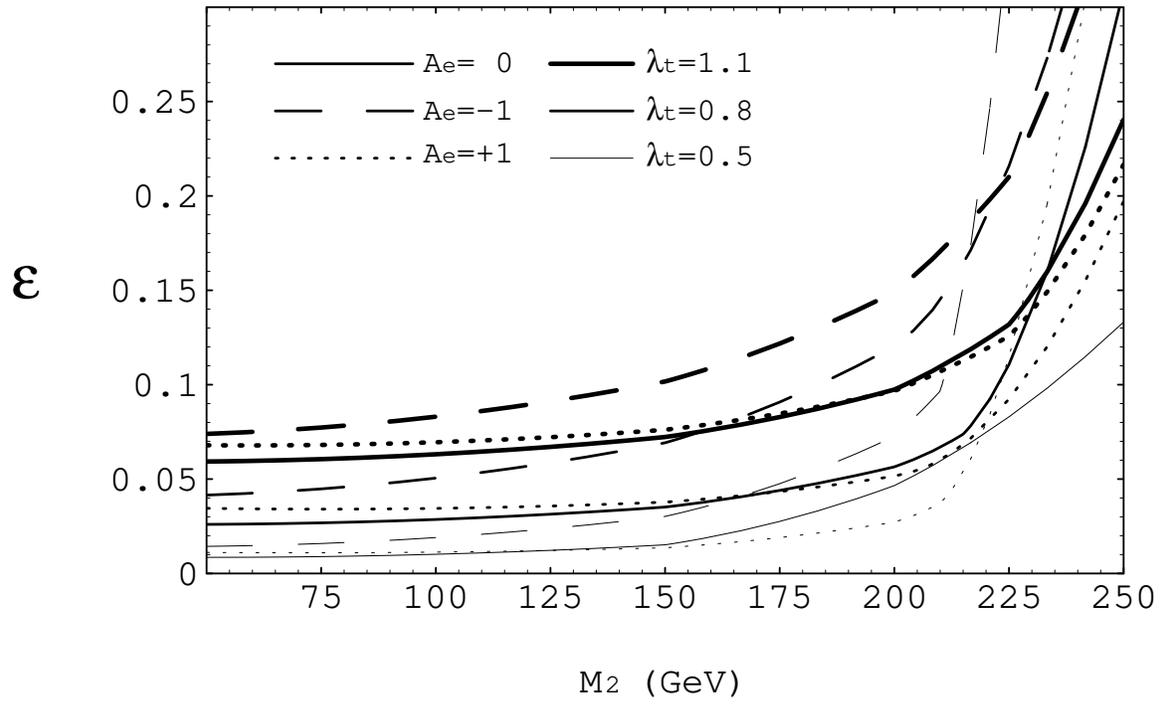
\item[Fig 10] Plots of the suppression factor $\epsilon$ against $M_2$, with
the same parameters as in Fig.\ 9.
\item[Fig. 11] Contours for $\Omega h^2$ in $\mu - M_2$ plane in the large
$\tan\beta$ limit. Dashed lines are LSP mass contours
of 30, 45, and 60 GeV.
For all regions of
$m_{LSP} < 45$ GeV in this plot, the Higgsino components of LSP are too big
and therefore they are ruled out by the $Z$ width.
\end{enumerate}
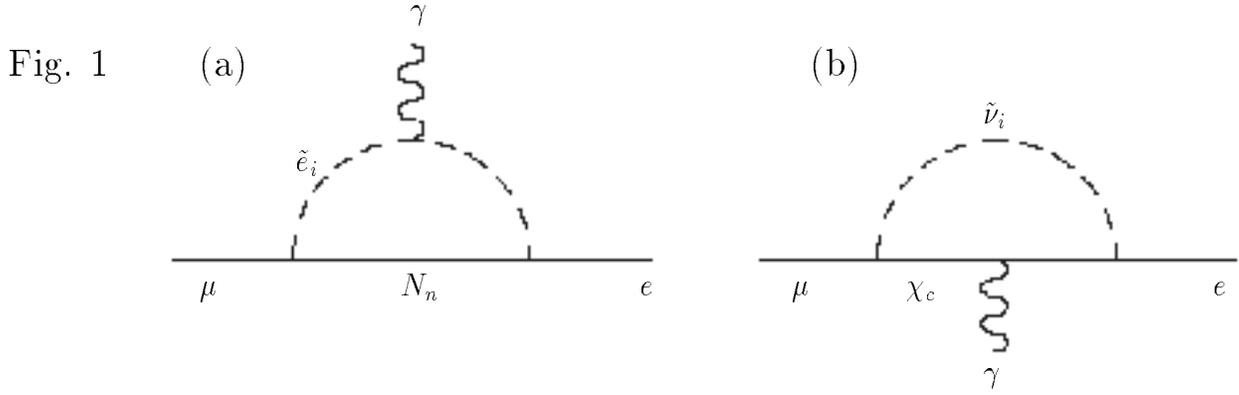
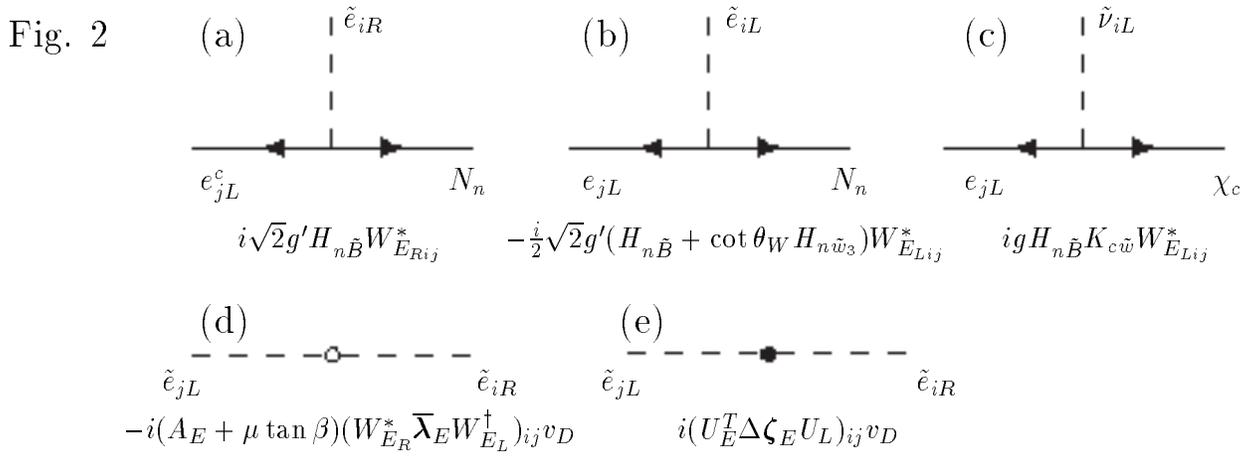
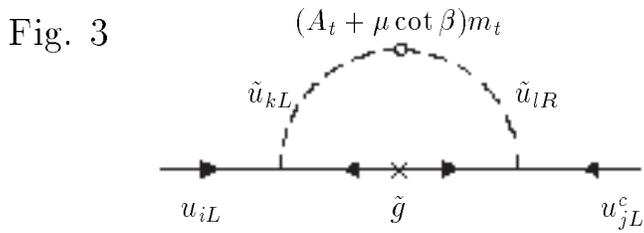
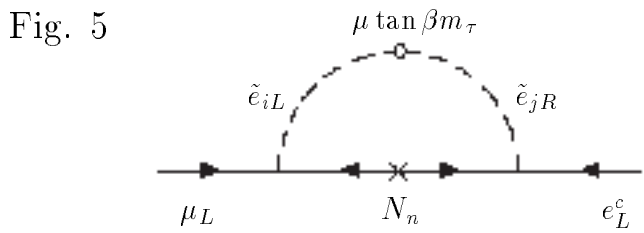
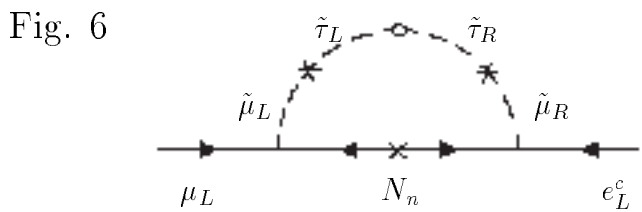
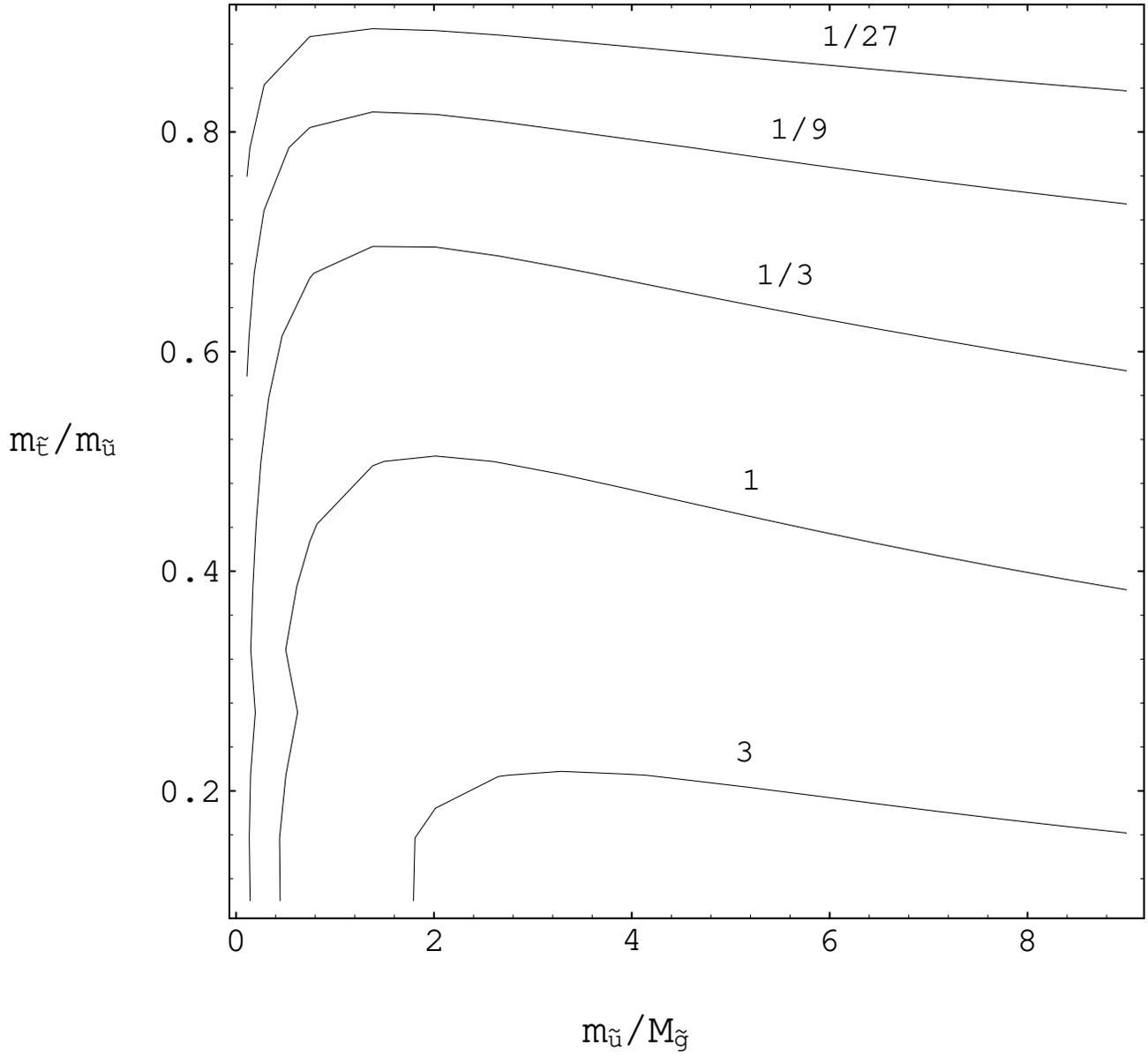
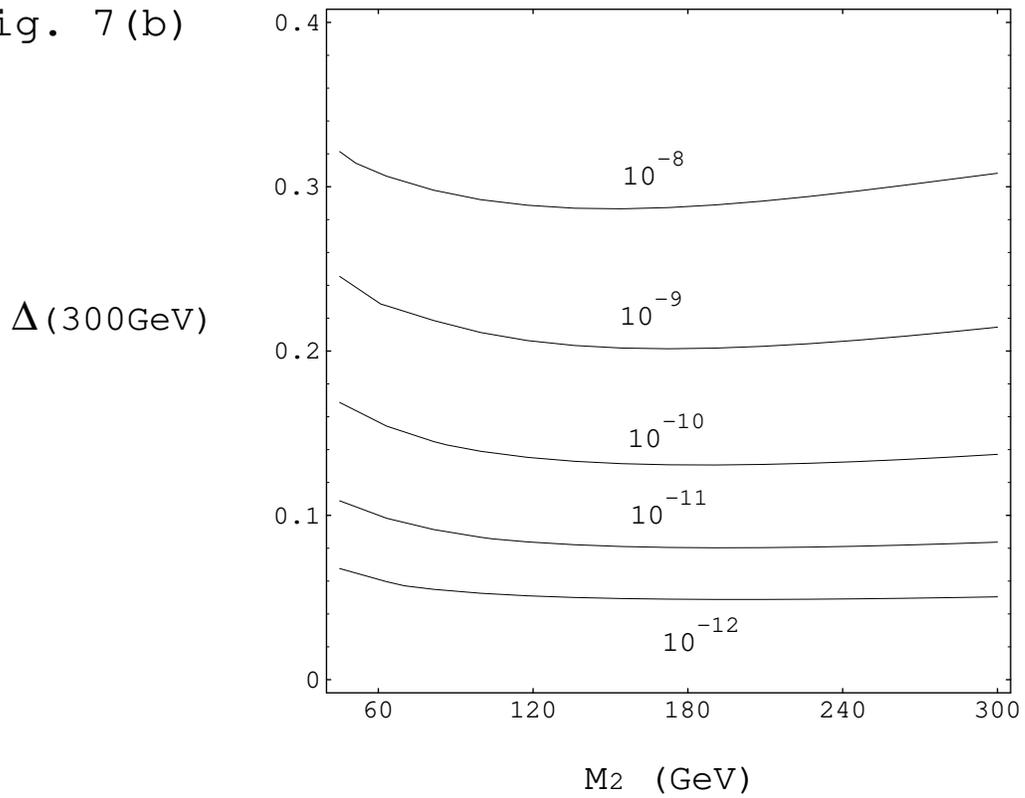
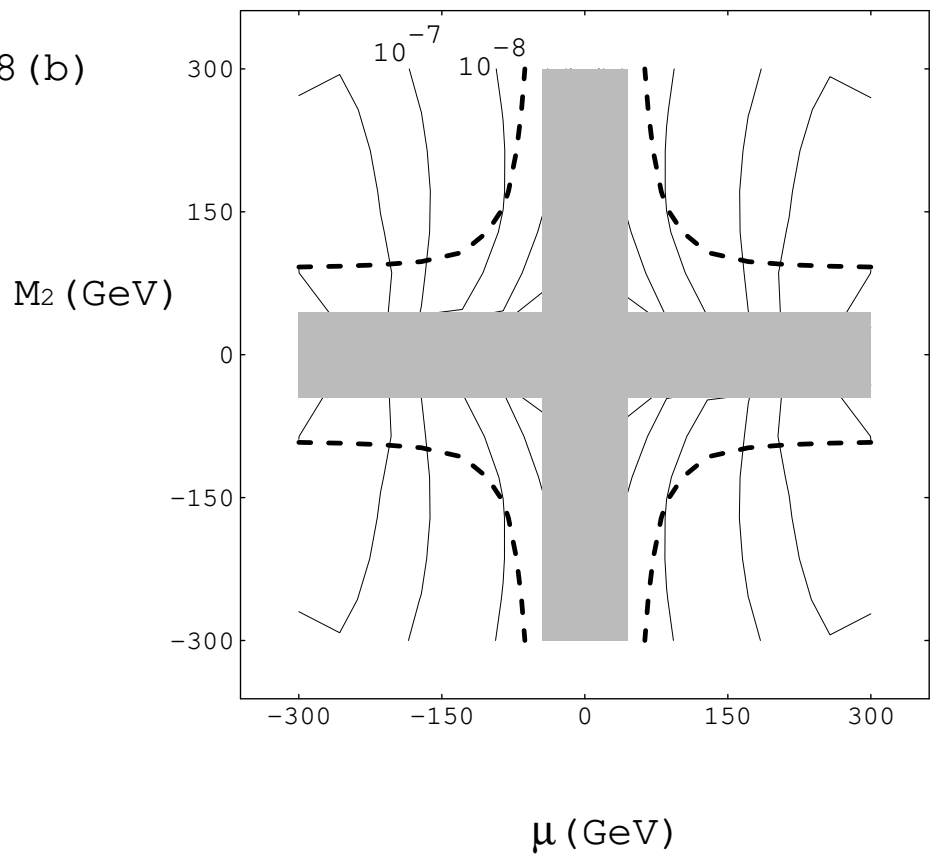
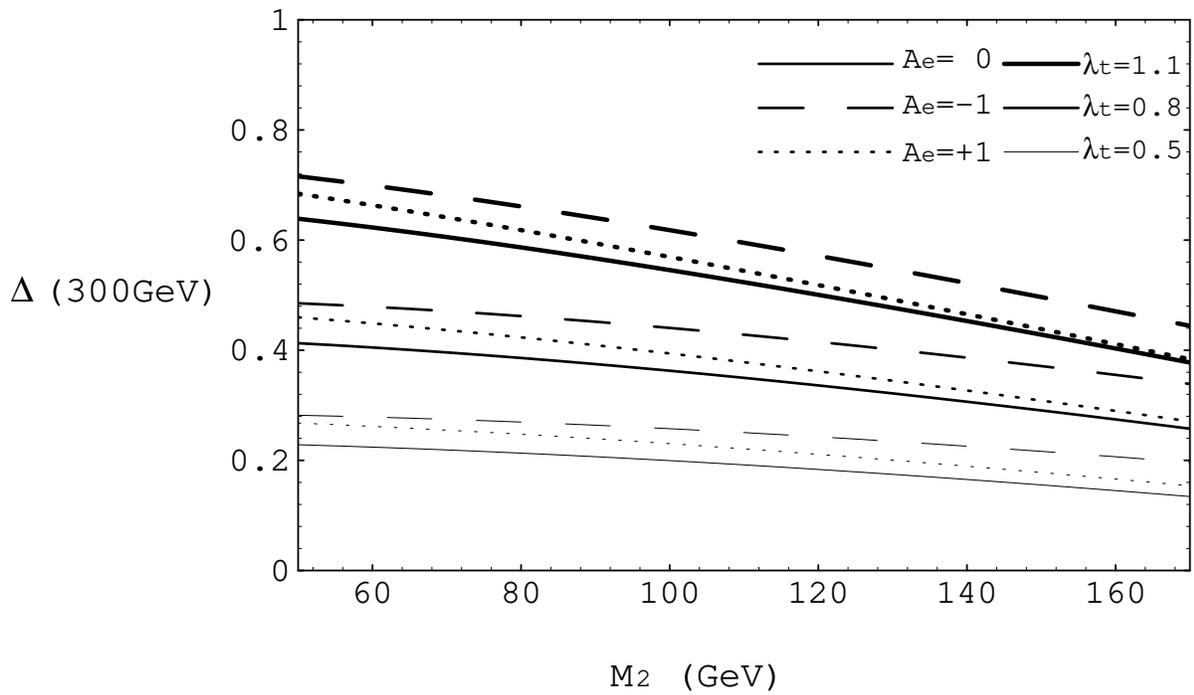
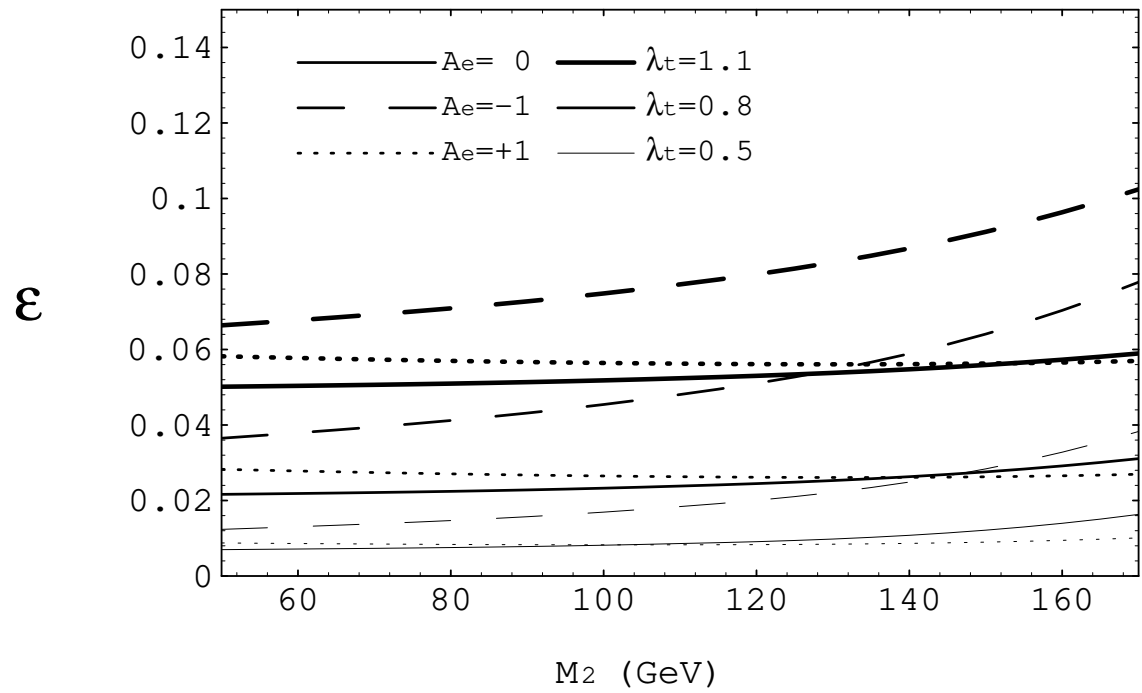
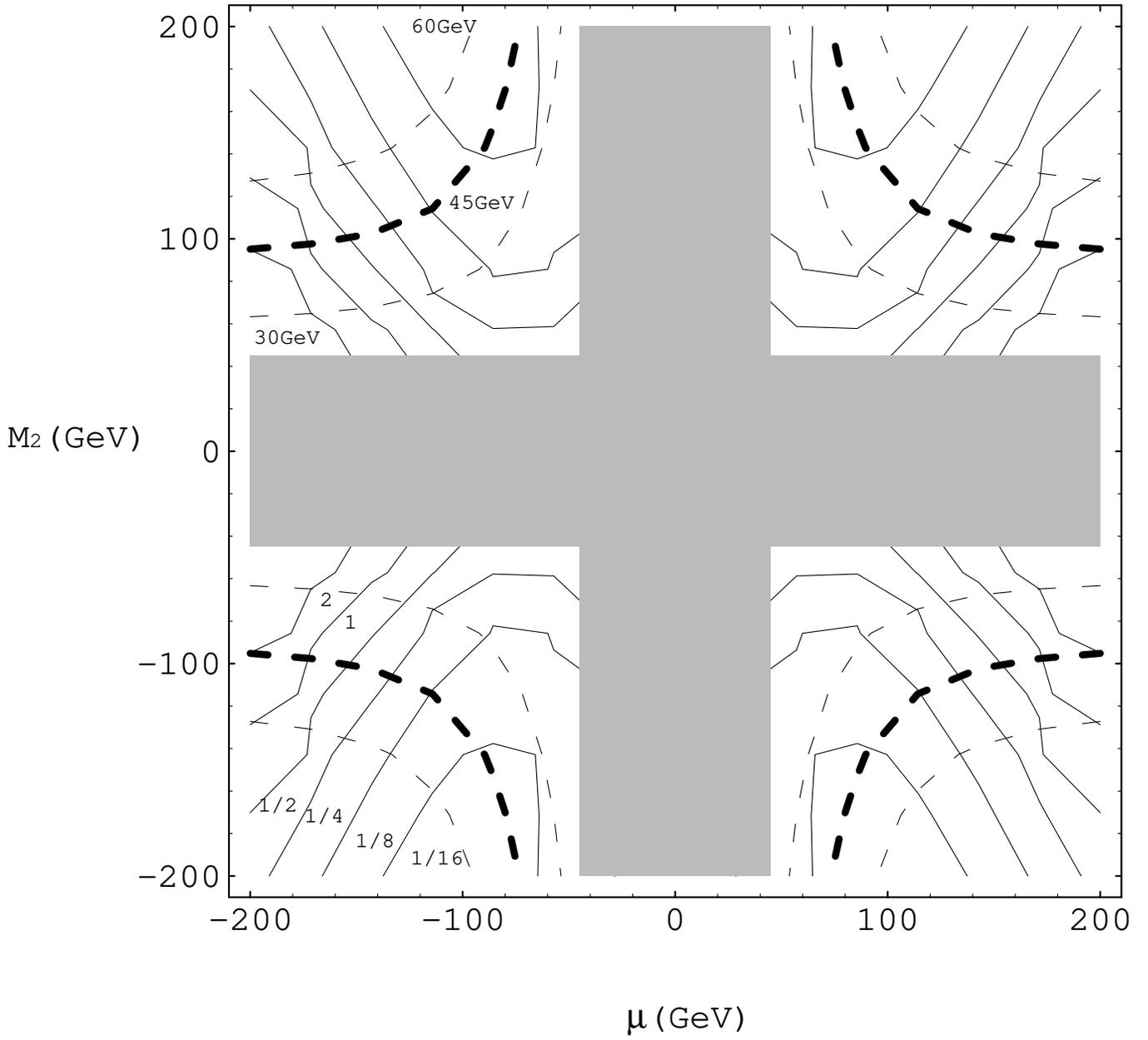
\end{document}